\documentclass[3p,preprint,12pt]{elsarticle}

\let\today\relax
\makeatletter
\def\ps@pprintTitle{%
    \let\@oddhead\@empty
    \let\@evenhead\@empty
    \def\@oddfoot{\footnotesize\itshape
         {Submitted preprint} \hfill\today}%
    \let\@evenfoot\@oddfoot
    }
\makeatother

    \usepackage{tgtermes}
    \usepackage{lineno}

    \usepackage{hhline}  
    \usepackage{float}

    \usepackage{multirow}
    \usepackage{graphicx,booktabs}

    \usepackage{xfrac}
    \usepackage[hidelinks]{hyperref}
    \usepackage{tabularx}

    \usepackage{wrapfig,booktabs}
    \usepackage{mwe}
    \usepackage{adjustbox}

    \usepackage{caption}
    \usepackage{subcaption}
    \usepackage{makecell}
    \usepackage{amsmath}
    \usepackage{amsmath,array}
    \usepackage[usenames, dvipsnames]{color}
    \usepackage[table]{xcolor}
    \usepackage{array}
    \usepackage{longtable}
    \usepackage{ragged2e}

    \usepackage{mathtools}
    \usepackage{setspace}
    \usepackage{tabu}
    \usepackage{natbib}
    \usepackage{amsmath}

    \usepackage{MnSymbol}
    \usepackage{xcolor}
    \usepackage{algpseudocode}
    \usepackage{soul}

    \usepackage{chngcntr}

    \newcommand{\noop}[1]{}
    \vspace{5mm}

    \biboptions{numbers,sort&compress}

\usepackage{multirow}

\begin{document}

\begin{frontmatter}

\title{High-throughput Alloy and Process Design for Metal Additive Manufacturing}

\author{Sofia Sheikh$^{a}$, Brent Vela$^{a}$, Pejman Honarmandi$^{a}$, Peter Morcos$^{a}$, David Shoukr$^{c}$, Abdelrahman Mostafa Kotb$^c$, Ibrahim Karaman$^{a,b}$, Alaa Elwany$^{a,c}$, Raymundo Arr\'{o}yave$^{a,b,c}$}
\address{$^a$Department of Materials Science and Engineering, Texas A\&M University, College Station, TX, USA}
\address{$^b$Department of Mechanical Engineering, Texas A\&M University, College Station, Texas, USA}
\address{$^c$Department of Industrial and Systems Engineering, Texas A\&M University, College Station, Texas, USA}

\cortext[mycorrespondingauthor]{Corresponding author email:\textrm{sofiasheikh@tamu.edu}}

\begin{abstract}

Designing alloys for additive manufacturing (AM) presents significant opportunities. Still, the chemical composition and processing conditions required for printability (ie., their suitability for fabrication via AM) are challenging to explore using solely experimental means. In this work, we develop a high-throughput (HTP) computational framework to guide the search for highly printable alloys and appropriate processing parameters. The framework uses material properties from state-of-the-art databases, processing parameters, and simulated melt pool profiles to predict process-induced defects, such as lack-of-fusion, keyholing, and balling. We accelerate the printability assessment using a deep learning surrogate for a thermal model, enabling a 1,000-fold acceleration in assessing the printability of a given alloy at no loss in accuracy when compared with conventional physics-based thermal models. We verify and validate the framework by constructing printability maps for the CoCrFeMnNi Cantor alloy system and comparing our predictions to an exhaustive 'in-house' database. The framework enables the systematic investigation of the printability of a wide range of alloys in the broader Co-Cr-Fe-Mn-Ni HEA system. We identified the most promising alloys that were suitable for high-temperature applications and had the narrowest solidification ranges, and that was the least susceptible to balling, hot-cracking, and the formation of macroscopic printing defects. A new metric for the global printability of an alloy is constructed and is further used for the ranking of candidate alloys. The proposed framework is expected to be integrated into ICME approaches to accelerate the discovery and optimization of novel high-performance, printable alloys.
\end{abstract}


\end{frontmatter}
\section{Introduction}
\label{sec:Intro}

Additive manufacturing (AM) is a growing technology in manufacturing industrial parts due to its capability of producing precise, complex shapes with high speeds through computer-assisted layer-by-layer part construction\cite{gardan2017additive}. In the case of metals AM, there is currently a broad, ever-expanding technology landscape with varying degrees of adoption in industry\cite{vafadar2021advances}. Among the many AM technologies presently available, liquid-mediated metal AM technologies have received widespread interest from academia and industry. For example, in the case of Laser Powder Bed Fusion (LPBF), the quality of the final manufactured product requires a thorough understanding of the interplay between process parameters, materials properties, and solidification conditions. Significant effort has been invested in understanding, for a given alloy system, the critical process conditions responsible for controlling the outcome of the fabrication process. Given the expense of powder fabrication and the cost to print a part, there is a need to be confident in the window of processing parameters that result in defect-free prints for a given alloy.

To date, the vast majority of work on metal AM has been framed regarding the need to tune processing conditions for the AM of specific alloys, typically developed for different manufacturing/fabrication routes  \cite{LIU2019107552,KOTADIA2021102155,ADEYEMI201818510}. This approach often overlooks that, historically, every engineering alloy system has always been designed with a processing route in mind. Historically, the early stages of research on metal AM focused on identifying process windows capable of yielding AM of conventional engineering alloys with properties comparable to (or even better than) their ingot metallurgy wrought counterparts. 

Under traditional alloy design schemes, the optimization of alloy performance tends to be carried out sequentially. After an alloy has been optimized against a handful of objective requirements, the focus tends to be shifted toward identifying a suitable scalable synthesis, manufacturing, and/or processing protocol to translate a material capability into a manufactured part or component that meets a specific technological need. Recently, it has been recognized that the underlying paradigm for metal AM must be shifted towards one in which alloys need to be designed specifically for AM  \cite{BANDYOPADHYAY2022207,ACKERS2021158965}. However, given the tight coupling between process and chemistry effects on resulting AM microstructures, the paradigm must instead shift to one in which alloy chemistry and processing windows are simultaneously designed for a specific application to minimize the presence of micro/macroscopic defects, thus enhancing the reliability of the fabrication process.

It should be noted that the major (macroscopic) defects in LPBF AM include: lack-of-fusion resulting from either the lack of powder melting (the lack of energy input) in the previous layer or the lack of overlap between the adjacent tracks or layers due to large hatch spacing or layer thickness~\cite{zhu2021predictive,zhang2017defect}; balling as a result of the insufficient energy input leading to the lack of molten liquid with high surface tension that reduces the wettability of the liquid on the substrate~\cite{zhu2021predictive,nasab2018morphological}; keyholing occurring at the excessive energy input resulting in porosities due to the high elemental evaporation~\cite{zhu2021predictive,forien2020detecting} and a deep liquid penetration due to the evaporation-induced recoil pressure on the melt pool surface~\cite{HONARMANDI2021102300,king2014observation}; cracking originating from the combination of high thermal gradient and residual stresses~\cite{zhu2021predictive,zhang2017defect}; hot tearing (cracking) as a material-dependent defect induced by the tension associated with obstructed thermal contraction (shrinkage)~\cite{KOU2015366}. 

Given the tight coupling between thermal histories, thermo-physical properties, and defect-inducing physical phenomena, different combinations of process parameters are likely to result in a greater or smaller probability for the onset of a specific printing defect. In this regard, printability maps\cite{JOHNSON2019199} must be constructed for highly printable material systems to identify regions associated with the defects in the process space and subsequently determine the defect-free region for the acceleration of the AM product design. To date, it has already been shown how experiments, simulations, and machine learning (ML) approaches can be combined to assess the printability map for a given alloy chemistry~\cite{zhang2021efficient,SEEDE2020199,zhang2021printability,zhang2021predictive,zhang2022hybrid}. 

A significant issue, however, is how to translate the concept of printability assessment as a tool to explore vast alloy spaces. The considerable cost associated with manufacturing AM powder feedstock makes the high-throughput (\emph{HTP}) determination of the printability map of a large number of alloys exceedingly expensive, mainly if such approaches rely on experiments.  To rapidly construct these printability maps for high-throughput design exploration purposes, a fully computational framework must predict both \emph{intrinsic} (due to their thermo-physical properties) and \emph{extrinsic} (induced by process parameters and thermal boundary conditions) susceptibility to the formation of defects associated with alloys of arbitrary compositions. Such a framework can accelerate the AM design process by identifying both intrinsically printable alloys and their corresponding optimal processing windows. 

A key challenge is developing appropriate defect criteria to differentiate different defect regions in both the composition and process spaces. To address this challenge, in this work, alloy design criteria for intrinsic susceptibility to balling and hot cracking are deployed to guide design in the \emph{composition space}. Once candidate alloy chemistries are identified, criteria based on materials properties, processing parameters, and/or thermal history proposed in the literature~\cite{king2014observation,xue2021controlling,SEEDE2020199,zhang2021efficient,gan2021universal,zhu2021predictive} are applied to find the defect (lack-of-fusion, balling, keyholing) and defect-free regions in the \emph{process space}. 

In this work, we demonstrate the co-design of alloy chemistry and AM processing parameters through the use of an exclusively computational approach. Our framework combines fast-acting models for the thermal history of a melt pool under arbitrary processing conditions, CALPHAD-based models for the phase stability and thermo-physical properties of alloys, as well as different criteria for the onset of both intrinsic (e.g. hot-cracking susceptibility) and extrinsic defects (keyholing, lack-of-fusion). Our framework is sufficiently fast to be deployed in a high-throughput fashion. Given its considerable interest by the scientific community, we demonstrate the framework over the 'Cantor' high-entropy alloy space. In order to establish the validity of the framework, we validate our predictive printability assessment with experimental data in the literature on the stoichiometric Cantor alloy, FeNiCoCrMn. We introduce a metric, \emph{printability index} that can be used to assess how printable a given alloy is over a wide range of processing conditions. The framework is then employed to identify promising regions in this fcc HEA space with the lowest susceptibility for fabrication defects. To our knowledge, this is the first example of the HTP exploration of large compositionally complex alloy spaces in order to assess their printability.

\section{Assessing the Printability of Alloys}
\label{sec:printability}

The construction of printability maps for a given alloy relies on the use of different criteria connecting characteristics of the melt pool (i.e. melt pool geometry), powder bed (i.e. powder layer thickness), and thermal histories (i.e. thermal gradients, cooling rates). These criteria can then be used to characterize different regions in the process space most likely to produce certain types of defects. The region free of such defects constitutes the \emph{printable region} for a given alloy chemistry. Regarding processing-induced defect criteria, some of these criteria require information about the thermal history and melt pool geometry, emphasizing the need for a fast-acting thermal model with sufficient precision in the context of high-throughput design exploration. Recently, different thermal models proposed in the field of welding have been applied to AM processes. One popular thermal model is the Eagar-Tsai (E-T) model~\cite{eagar1983temperature}, which has been used to analyze defect criteria for the construction of printability maps. The validity of the E-T model in AM has been studied in many works, e.g.,~\cite{HONARMANDI2021102300, JOHNSON2019199, SEEDE2020199, xue2021controlling, zhang2021efficient, tapia2018uncertainty}. Among these works, Tapia et al.~\cite{tapia2018uncertainty} applied a general polynomial chaos expansion to propagate uncertainties from the input variables of the E-T model to its output, i.e., the melt pool geometry, and validate the probabilistic results with their corresponding experimental data for the 430F stainless steel. Johnson et al.~\cite{JOHNSON2019199} also showed that the melt pool dimensions obtained from the E-T model and a high-fidelity finite element (FE) COMSOL model exhibit similar errors for the experimental measurements on the Ni-5wt.\%Nb alloy, except for considerably higher discrepancies in the melt pool depth predicted by the E-T model at the keyholing mode due to the model’s missing physics for this mode. In 2021, Honarmandi et al.~\cite{HONARMANDI2021102300} probabilistically benchmarked the E-T model with clean, extensive experimental data for different materials systems. Large errors associated with the keyhole depth are corrected in this work using a simplified model proposed by Gladush and Smurov~\cite{gladush2011physics} by solving the heat conduction equation in the case of a semi-infinite uniform substrate and a cylindrically-shaped melt pool. 

Due to the overall validity of the E-T model, recent works in literature have taken advantage of this thermal model to produce printability maps for various material systems such as Ni-5wt.\%Nb alloy~\cite{JOHNSON2019199}, Co-Cr-Fe-Mn-Ni high entropy alloy (HEA)~\cite{JOHNSON2019199, zhang2021efficient}, AF9628 low alloy martensitic steel~\cite{SEEDE2020199, zhang2021efficient}, and Ni-Ti alloys~\cite{xue2021controlling}, printed via laser powder bed fusion (L-PBF). In a newly published article, Islam et al.~\cite{islam2022high} also used a simplified analytical thermal model to construct the printability map for a Ni-based superalloy. However, in the aforementioned works, many L-PBF experiments across the process parameter space were performed to calibrate the thermal model parameters and determine criteria associated with various defect modes. These experiments create a bottleneck in AM design process and restrict HTP AM design approaches. In the present work, the expensive calibration step is avoided for the sake of high-throughput design by estimating the materials’ properties through the CALPHAD models in the ThermoCalc software.

To further accelerate HTP printability analyses, some universal scaling laws in terms of dimensionless numbers have been recently proposed to reduce printability analyses to spreadsheet calculations~\cite{rubenchik2018scaling,rankouhi2021dimensionless}:

Islam et al.~\cite{islam2022high} applied a dimensionless number proposed by Rankouhi et al.~\cite{rankouhi2021dimensionless}---i.e., $\Pi_1=\frac{C_p P}{k \nu^2 h}$ where $P$, $\nu$, $C_p$, $k$, and $h$ are the heat source power, heat source speed, specific heat, thermal conductivity, and hatch spacing, respectively--- and a scaling law for the relative density in terms of $\Pi_1$~\cite{rankouhi2021dimensionless,islam2022high}. This scaling law was validated with HTP experiments conducted to differentiate the printable (defect-free) region from the defect regions for various values of $\Pi_1$. While this work is fully computational, material properties were adopted from literature for a nickel-based superalloy which restricts this framework to materials for which there is experimental data available. This framework is only able to identify the lack-of-fusion and keyholing regions in the process space based on the porosity-induced density changes but is not able to detect the balling defect.

In another recent work, Zhu et al.~\cite{zhu2021predictive} applied the scaling law introduced by Rubenchik et al.~\cite{rubenchik2018scaling} to predict melt pool dimensions during the L-PBF AM processes as functions of two dimensionless variables---i.e., $p=\frac{D}{a \nu}$ and $B=\frac{\Delta H}{2^{(3/4)} \pi h_s}$ where $D$, $a$, and $\frac{\Delta H}{h_s}$ are the thermal diffusivity, the laser beam radius, and the normalized enthalpy, respectively---faster than the E-T model. This work is also fully computational and is the closest to the framework proposed in the present work for the construction of printability maps. However, in the present work we used and compared three different thermal models for the prediction of melt pool geometry and twelve different combinations of defect criteria for the generation of printability maps for an entire alloy system, i.e. Cantor alloys, as opposed to only using the aforementioned scaled thermal model with one set of defect criteria for a specific material, i.e., Ni-Ti shape memory alloy (SMA), as presented in Zhu et al.'s work~\cite{zhu2021predictive}. 

Another approach to accelerate printability map production for the sake of high-throughput AM design is the application of machine learning (ML) models. Scime and Beuth~\cite{scime2019using} applied a combination of feature extraction methods and unsupervised ML approaches to detect different melt pool morphologies corresponding to different types of print defects. They indicated their prevalence across the process space for the Inconel 718 alloy. Du et al.~\cite{du2021physics} implemented a physics-informed ML model trained against binary data, i.e., “balling” or “no balling” occurrences, for six different alloys to predict balling susceptibility indices and maps in terms of six relevant mechanistic variables. As can be noticed, these models have been constructed for specific material systems with few training data points. In our work, a neural network (NN) model with many relevant material properties (i.e. thermal conductivity, specific heat, etc.) is trained using a large and validated computational database for different material systems to connect the chemistry and processing conditions to melt pool geometry during the AM processes in a material-agnostic manner. These relevant material properties are used to estimate the melt pool dimensions of arbitrary alloys. These melt pool dimensions are then used to construct printability maps based on the defect criteria. 

Akbari et al.’s~\cite{akbari2022meltpoolnet} leveraged an ML regression and classification package called MeltpoolNet to train models to both predict melt pool geometry and defect modes. This model was trained on a wide range of experimental data collected for various material systems. However, it seems their results for printability maps show insufficient predictability compared to the experimentally-derived printability maps in the literature. This can probably be attributed to the sparsity of training data used in this work and the well-known inability of NNs to make robust predictions in the sparse data regime. 

As mentioned, a computational package is developed in this work to accelerate the construction of printability maps for the sake of the high-throughput design of defect-free AM parts with desirable properties and performance. This package includes three thermal models, i.e., the E-T model, the scaled E-T model, and the NN model, in addition to different defect criteria, i.e., two criteria for lack-of-fusion, two criteria for balling, and three criteria for keyholing, to identify the defect and defect-free regions in the printability maps. The printability maps, obtained from three different models and twelve combinations of defect criteria, are analyzed and compared for the FeCoCrMnNi equiatomic Cantor alloy. This alloy is selected as a case study as it is one of the most well-studied HEAs due to its excellent fracture toughness and ductility~\cite{zeng2021mechanical}. After identifying the optimal criteria set, the printability maps of five compositions in the same material system that show the least susceptibility to balling and hot cracking defects are constructed in order to co-design the most favorable alloy compositions and processing conditions toward the defect-free AM parts. We believe the framework proposed in this work indicates a promising avenue for accelerated material and process design in AM.

\section{Prior Work on  the Design of High Entropy Alloys for Laser Powder-Bed Fusion}

High entropy alloys (HEAs) have attracted an increasing interest in the past few years due to their outstanding mechanical properties ~\cite{ref1,paper1,paper2,paper31,paper33,paper58,paper84,paper85,paper89,paper91,paper99,paper101,paper102,paper103,paper104,paper106,paper107,ref2,paper118,paper124,paper126,paper127,paper130,paper131,paper141}, high-temperature mechanical properties~\cite{paper260} and resistance to corrosion and oxidation~\cite{paper1,ref3}. High entropy alloys consist of five or more principal elements at a concentration of 5 at\%, minimum. To date, the few systems that have been widely researched include the equiatomic CoCrFeMnNi, CoCrFeNi, Al$_x$CoCrFeNi, where $x$ is the molar ratio, and a limited number of refractory HEA systems. Various trial-and-error strategies have been used to print HEAs. The effect of power, velocity, hatch spacing, laser beam diameter, and powder layer thickness have been studied to optimize the processing conditions for HEAs ~\cite{paper1,paper31,paper60,paper61,paper85,paper86,paper87,paper96,paper107,paper117,paper127,paper130,paper105}. To optimize for processing parameters of built samples, the Archimedes density has been used as a design \emph{constraint} while the volume energy density (VED) is used as a design \emph{objective} to optimize the printability of an alloy system ~\cite{paper31,paper33,paper86,paper87,paper91,paper99,paper102,ref2,paper122,paper141}. For example, for the CoCrFeMnNi system, Dovgyy et al.~\cite{paper141} was able to fabricate a sample with a relative density greater than 99\% at a VED value of 62.7 J/mm$^3$. However, Dovgyy et al.~\cite{paper141} also showed that the density fluctuated at the sample energy density. Likewise, Kim et al.~\cite{paper118} was able to fabricate CoCrFeMnNi system with greater than 99\% density at a VED value of 75 $J/mm^3$. 

The solidification microstructure and mechanical properties have been optimized by various techniques. To optimize for the mechanical properties of the as-built sample, interstitial atoms such as carbon and nitrogen were added to the powder \cite{paper103,paper88,paper127,paper104,paper100}. The addition of the interstitial atoms acts as precipitates that impede the movement of dislocations during plastic deformation, improving the mechanical properties of the system. Park et al.~\cite{paper127} doped CoCrFeMnNi with 1\% carbon and obtained a higher strength value compared to as-built CoCrFeMnNi HEAs. In addition to interstitial atoms, nano-TiN particles were used to enhance the printability and mechanical behavior of CoCrFeNiMn alloys~\cite{paper122}. In this regard, Li et al.~\cite{paper122} mixed nano-TiN particles with the HEA powder to achieve higher mechanical properties than as-built samples. 

Other than trial-and-error methods, processing maps were proposed by Chen et al.~\cite{paper102} to fabricate a build with high density. Johnson et al.~\cite{JOHNSON2019199} also proposed a methodology to predict the printability of an alloy subjected to L-PBF where the regions in the process space which result in keyholing, balling, and lack-of-fusion are identified. Such identification is known as a printability map. The geometry of the melt pool used to predict the defect-mode boundaries was predicted using a finite element model (FEM). The printability maps generated using the high-fidelity FEM were compared to those that were predicted using a simpler thermal model, the Eagar-Tsai model. Furthermore, the uncertainty of the bounds for the different defect regions was quantified. However, Johnson et al.~\cite{JOHNSON2019199} only used one set of criteria for lack-of-fusion, balling, and keyholing to generate the printability maps. Additionally, the thermophysical properties required by the thermal models were obtained from experiments, which can be an expensive process. In addition, the prevailing issue of hot cracking in HEAs was not addressed.

HEAs are susceptible to hot crack during L-PBF~\cite{ref1,ref2,paper30,paper32,paper96,paper129, paper131}. Cracking in alloys such as the CoCrFeMnNi system is due to the thermal stresses the material undergoes due to the rapid solidification during L-PBF. Due to hot cracking, the properties and performance of a material can deteriorate. A few design strategies have been proposed to remedy hot cracking and obtain a denser build.  One method proposed is to heat the substrate of the build to reduce the thermal stress experienced by the material during printing ~\cite{paper60,paper96}. Yao et. al.~\cite{paper60} used an orthogonal design of experiments with a substrate heated above room temperature to find an optimal combination of processing parameters for an aluminum-based HEA. 

Studying the effect of the scan strategies on the build is another method that has been employed to remedy the effect of thermal stresses to eliminate hot-cracking ~\cite{paper141,paper131,paper124,paper88}. Zhang et al.~\cite{paper131} suggested rotating the build layers along the X-axis at three different angles in an attempt to control the distribution of thermal stress exhibited by the CoCrFeMnNi sample during the L-PBF process and thereby reduce crack sensitivity. However, hot cracking was still prevalent in the samples. Thermo-mechanical processing on the as-built sample was another methodology proposed to help with the hot cracking problem in the materials. Various thermo-mechanical processes, such as annealing~\cite{paper106,paper97,paper99} and hot isostatic pressing~\cite{paper86}, have been proposed to increase the densification of the print as well as remediate hot-cracking. 

\begin{figure}[htb!]
    \centering
    \includegraphics[width = \linewidth]{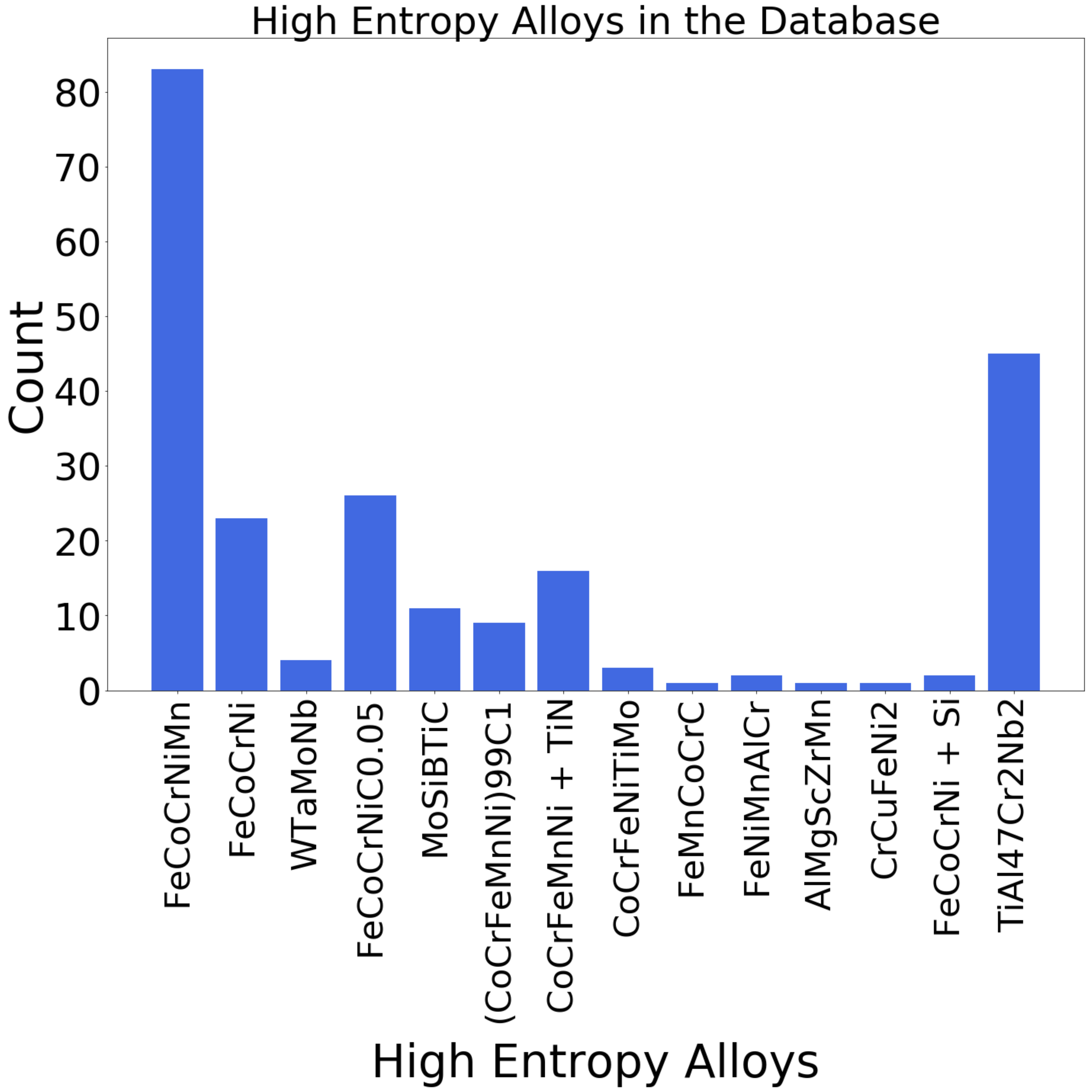}
    \caption{Number of studies on different HEAs in the literature. It shows that HEAs with the element Co, Cr, Fe, Mn, and Ni are one of the most widely researched alloys and therefore, we can narrow down our composition space such that we only consider HEAs with the suggested elements.}
    \label{fig:HEAs_Data_Sh}
\end{figure}

The Cantor alloy system (Co-Cr-Fe-Mn-Ni) is one of the most well-studied HEA systems, both theoretically and experimentally, as seen in Figure \ref{fig:HEAs_Data_Sh}. Among these alloys, equiatomic CoCrFeMnNi exhibits a single-phase face-centered cubic (FCC) phase structure that provides good mechanical properties \cite{ref2}. However, as mentioned above, hot cracking is still a prevalent issue for the printability of this class of alloys. Previous design of L-PBF experiments for high entropy alloys relies on high-fidelity simulations or trial-and-error experiments that are expensive and time-consuming. The methodology proposed in this work can reduce these costs by intelligently allocating experimental resources to study alloys that are \emph{amenable} to L-PBF. In this work, we design Cantor alloys that are resistant to hot cracking and balling and identify the window of processing parameters that result in defect-free prints for these printable alloys.

\section{Model Description}\label{sec:model_des}
\subsection{Calculation of Thermophysical Properties}\label{sec:thermal_prop}
Several models are implemented to calculate the thermo-physical properties across the Cantor alloy system. Thermo-physical properties are derived using the Property, Equilibrium, and Scheil CALPHAD-models in the Thermo-Calc software using thermodynamic databases: TCHEA5, TCAL7, and TCNI11. Various material properties relevant to the L-PBF process are calculated, such as the solidus and liquidus temperatures, the latent heat of fusion, etc. Furthermore, at room temperature (298 K), solidus temperature, liquidus temperature, the following material properties are queried from the software: density, electric conductivity, heat capacity, thermal conductivity, thermal diffusivity, thermal resistivity, enthalpy, dynamic viscosity, and kinematic viscosity. Besides CALPHAD–based models,  reduced-order models (ROMs), $P_{alloy} = \Sigma_{i}^n x_{i}P_{i}$---where $i$ is the set of atomic species, $x$ is the composition value and $P$ is the material property---are used to calculate (the average) melting temperature, boiling temperature, density, and molecular weight. Here, the elemental properties are obtained from Pymatgen (Python Materials Genomics), a Python library. The CALPHAD–derived and ROM-derived properties are used as inputs to thermal models and variables in AM–related dimensionless numbers, as discussed in Section \ref{sec:thermal_models}. 

Additional properties are calculated using the CALPHAD-based material properties and ROM-calculated properties. For example, the solidification range can be calculated as $T_{liquidus} - T_{solidus}$. Moreover, the total enthalpy can be defined as $H_{liquidus} - H_{RT}$, where $H_{liquidus}$ is the enthalpy at liquidus and $H_{RT}$ is the enthalpy at room temperature. Using the total enthalpy, the effective heat capacity can be calculated as $H_{total}/(T_{liquidus} - T_{RT})$. The melting enthalpy and boiling enthalpy were quantified using $H_{liquidus} - H_{solidus}$, and $10 \times R \times T_{liquidus}$, respectively. H$_{solidus}$ is the enthalpy at the solidus temperature and $R$ is the gas constant. Furthermore, the enthalpy at boiling and enthalpy after boiling was calculated using $H_{liquidus} + C_{p,liquidus} \times (T_{boiling} - T_{liquidus})$ and $H_{at-boiling} + H_{boiling}$, where $C_{p,liquidus}$ is the specific heat at the liquidus temperature, $T_{boiling}$ is the boiling temperature, $H_{at-boiling}$ is the enthalpy at boiling and $H_{boiling}$ is the boiling enthalpy. 

Furthermore, a composition-based feature vector (CBFV) is calculated to gain further insight into the material system. A CBFV uses a data-driven approach to calculate physically interpretable results about the material system~\cite{murdock2020domain}. The construction of CBFV has been used in materials informatics studies and contains descriptive statistics such as the average, range, sum, and variance of the constituent elements. In the package, the alloy system is featurized using the Oliynik data set, which uses element chemistry to predict the stoichiometric material properties~\cite{oliynyk2016high}. 

\subsection{Thermal Model}\label{sec:thermal_models}

\subsubsection{Eagar-Tsai Model}

In 1983, T. W. Eagar and N. S. Tsai introduced the Eagar-Tsai model (E-T model)~\cite{eagar1983temperature}. E-T model is an inexpensive analytical model that can predict the temperature distribution for a moving heat source with a Gaussian profile on a substrate during the welding process. Overall, good agreement between the calculated melt pool dimensions from the E-T model and the experimental data has made the E-T model a reliable method for calculating the melt pool dimensions of single tracks within the AM community~\cite{SEEDE2020199, Promoppatum2019, JOHNSON2019199,  rubenchik2018scaling}. The temperature profile of the substrate is expressed by the Green’s function for the steady-state heat conduction equation. Due to the complexity of the solution in the presence of a double integral, the solution is simplified by approximating the input heat source profile with a Gaussian function. Additionally, a quasi-steady state semi-infinite substrate with temperature-independent material properties is considered, while the convective and radiative heat flow is ignored to simplify the model. The final form of the integration used to calculate the temperature distribution is expressed in the following form:

\begin{equation}
\label{ET}
\begin{split}
T - T_0 = \frac{1}{2} \int_{t' = 0}^{t' = t} {dT_{t'}} = \frac{q} {\pi \rho C_p \sqrt{4 \pi \alpha}} \\
\int_{t' = 0}^{t' = t} \Big[ \frac{(t - t')^{-\frac{1}{2}}}{2 \alpha (t - t') + \sigma^2}
e^{[-\frac{(x - \nu t')^2 + y^2}{4 \alpha (t - t') + 2 \sigma^2} - \frac{z^2}{4 \alpha (t - t')}]} \Big] dt',
\end{split}
\end{equation}

In the above equation, $T_{0}$ is the initial temperature of the substrate, $T$ is the predicted temperature at time \textit{t} at any point with coordinate of (\textit{x,y,x}). The input variables for the E-T model fall into two categories: \emph{process parameters} and \emph{material properties}. The process parameters are $q$, $\mu$, and $\sigma$ which denote the input heat energy to the material calculated by multiplying the laser heat source power times the efficiency, the scanning speed of the moving heat source, and the standard deviation of the Gaussian laser profile, respectively. On the other hand, material properties include $\rho$ which is the density of the printed material, $C_{p}$ which is the specific heat capacity, and $\alpha$ which is the thermal diffusivity. The following equation calculates $\alpha$:

\begin{equation}
\label{thermaldiffusivity}
\alpha = \frac{\kappa}{\rho C_{p}} 
\end{equation}

In order to estimate the melt pool length, width, and depth, the temperature at each point is calculated by applying the numerical quadrature method to Equation~\ref{ET}. The boundary of the melt pool is then determined by comparing the calculated temperature to the melting temperature of the material~\cite{HONARMANDI2021102300}.

\subsubsection{Eagar-Tsai Model in Dimensionless Form}

For a faster and more convenient calculation of melt pool dimensions, the E-T model in Equation~\ref{ET} can be expressed in a dimensionless form by converting variables $x$, $y$, and $z$ into normalized dimensionless variables $x_n=\frac{x}{a}$, $y_n=\frac{y}{a}$, and $z_n=\frac{z}{\sqrt{D/(ua)}}$, respectively~\cite{rubenchik2018scaling,zhu2021predictive}. This change of variables results in the following solution for Equation~\ref{ET}:

\begin{equation}
\label{eq: E-T dimensionless solution}
T=T_s g(x_n,y_n,z_n,p)=T_m B g(x_n,y_n,z_n,p)    
\end{equation}

In the above equation, $g$ is a universal function in terms of the given coordinate and dimensionless parameter $p$ that is expressed as the laser dwell time ($t_d = \frac{a}{\nu}$) over the thermal diffusion time ($t_D = \frac{a^2}{D}$), i.e., $p=\frac{D}{ua}$. This function captures the effect of the laser scan velocity on the temperature distribution during the AM processes. $B$ is another dimensionless parameter defined as the energy density deposited by the laser beam divided by the enthalpy of melting ($h_s=\rho C T_m$), i.e., $B = \frac{\Delta H}{2^{3/4} \pi h_s}$; $B$ captures the effect of the laser power on the temperature distribution during L-PBF. 

It should be noted when $p$ is smaller than 1, the diffusion time is greater than the dwell time. In other words, the beam size is greater than the thermal diffusion into depth during the dwell time, leading to a shallow and extended melt pool. This situation typically takes place when the material thermal conductivity is low. On the other hand, a $p$ greater than 1 usually happens in materials with high thermal conductivity, resulting in a higher thermal diffusion into depth and a more circular shape for the melt pool. $B$ should also be much greater than 1 to have enough energy input to form a well-defined melt pool. 

The melt pool boundary must be determined along each normalized coordinate axis to estimate the normalized melt pool length, width, and depth. This is equivalent to finding the spatial point on each axis where the predicted temperature in Equation~\ref{eq: E-T dimensionless solution} equals the melting temperature ($T=T_m$), i.e.,:

\begin{equation}
\label{eq: Melt pool boundaries}
B g(x_n,y_n,z_n,p) = 1
\end{equation}

As a result, the melt pool dimensions can be predicted in terms of dimensionless variables $p$ and $B$ in a universal scheme. For this purpose, Rubenchik et al.~\cite{rubenchik2018scaling} interpolated the melt pool dimensions calculated for different combinations of processing parameters and materials systems. They showed that $p$ and $B$ usually change in the range of 0.1-5 and 1-20 in these calculations, respectively. These interpolated functions are: 

\begin{equation}
\label{eq: Melt pool length}
\begin{split}
L(p,B) = \frac{a}{p^2} \Big[0.0053 - 0.21p + 1.3p^2 + (-0.11-0.17B) P^2 \ln(p) \\
 + B(-0.0062 + 0.23P + 0.75p^2) \Big]
\end{split}
\end{equation}

\begin{equation}
\label{eq: Melt pool width}
\begin{split}
W(p,B) = \frac{a}{B p^3} & \Big[0.0021 - 0.047p + 0.34p^2 - 1.9p^3 - 0.33p^4 + \\ & B(0.00066 - 0.0070p - 0.00059p^2 + 2.8p^3 \\ & - 0.12p^4) + B^2(-0.00070 + 0.015p - 0.12p^2 + 0.59p^3 \\ & - 0.023p^4) + B^3(0.00001 - 0.00022p + \\ & 0.0020p^2 - 0.0085p^3 + 0.0014p^4) \Big]
\end{split}
\end{equation}

\begin{equation}
\label{eq: Melt pool depth}
\begin{split}
D(p,B) = \frac{a}{\sqrt{p}} & \Big[0.008 - 0.0048B - 0.047p - 0.099Bp + \\ & (0.32 + 0.015B)p\ln(p) + \\ & \ln(B)(0.0056 - 0.89p + 0.29p\ln(p)) \Big]
\end{split}
\end{equation}

These universal algebraic functions reduce the time and complexity associated with the calculations of the melt pool dimensions in the E-T model, making it amenable for use in automated HTP materials design frameworks. Further discussion about the dimensionless E-T model and the interpolated functions for the melt pool dimension estimation can be found in~\cite{rubenchik2018scaling}. 

\subsubsection{Machine Learning Model}
A high-fidelity surrogate for the analytical E-T model was developed via neural networks (NN). As discussed in Section \ref{sec:printability}, ML models have been employed to accelerate and automate manufacturing in AM. A popular choice of ML technique that has been employed is NN. However, limited data can result in overfitting and a lack of generalization of the NN models. 

The purpose of deploying a surrogate model for E-T is to improve the melt pool dimension prediction with minimum computation time. To deploy such a predictive model, the framework consists of integrating tools of physics-based analytical modeling and data-driven analysis and calibrating it using statistical models. The framework can be divided into three steps:

\begin{enumerate}
    \item Obtaining data needed to develop the model
    \item Training and testing the NN model and creating a surrogate model for all possible physical inputs.
    \item Calibrating the model using the Bayesian method
\end{enumerate}

To develop the ML surrogate model, 550,000 simulations were generated using the analytical E-T model. The parameters used for the simulations were sampled using the Latin hypercube sampling (LHS) design. Table \ref{table:LHS_surrgate} displays the data ranges used for the LHS design. 

\renewcommand{\arraystretch}{1.5}
\begin{table*}
    \centering
    \setlength\extrarowheight{5pt}
    \begin{tabularx}{0.9\textwidth} { 
      | >{\centering\arraybackslash}X 
      | >{\centering\arraybackslash}X |}
     \hline
     \textbf{Property} & \textbf{Range} \\
    \hline
    Laser Speed & 0.05-5 m/s\\
    \hline
    Laser Power & 10-500 W\\
    \hline
    Beam Diameter & 50-150 $\mu$m\\
    \hline
    Absorptivity & 0.05-1 \\
    \hline
    Melting Temperature & 676-3,695 $^{\circ}$K \\
     \hline
     Density & 1,730-19,300 Kg/$m^{3}$\\
     \hline
     Thermal Conductivity & 10-234 W/m/$^{\circ}$K \\
     \hline
     Specific Heat & 450-1,400 J/Kg/$^{\circ}$K \\
    \hline
    \end{tabularx}
    \caption{Data ranges for the LHS design used to develop the surrogate model}. 
    \label{table:LHS_surrgate}
\end{table*}

Several different ML techniques were tested to develop the surrogate model, such as Linear Regression, Partial Least Squares, Gradient Boosting, and NN. A 10 k-fold cross-validation method was used to compare the performance of the 4 models where the performance metric used to evaluate the performance was Mean Square Error (MSE) as shown in Equation \ref{eq:MSE}, where y$^{Ph}$ is the physical observation, y$^{S}$ is the NN model prediction and N$_{P}$ is the total number of simulations. The relative performance of each model is quantified using Equation \ref{eq:MSE} are displayed in Table \ref{table:MSE_perf}.

\begin{equation}
\label{eq:MSE}
    MSE = \frac{ \sum_{i=1}^{N_{P}} (y_{i}^{Ph} - y_{i}^{S})^{2}}{N_{P}}
\end{equation}

The NN model outperformed the other models as shown in Table \ref{table:MSE_perf}, so NN was chosen as the ML technique for the surrogate model. 

\renewcommand{\arraystretch}{1.5}
\begin{table*}
    \centering
    \setlength\extrarowheight{5pt}
    \begin{tabularx}{0.9\textwidth} { 
      | >{\centering\arraybackslash}X 
      | >{\centering\arraybackslash}X |}
     \hline
     \textbf{Method} & \textbf{Relative MSE} \\
    \hline
    Linear Regression & 14.59\\
    \hline
    Partial Least Squares & 14.60 \\
    \hline
    Gradient Boosting & 1.30 \\
    \hline
    Neural Network & 1 \\
    \hline
    \end{tabularx}
    \caption{For the preliminary testing to determine which ML technique was necessary, relative MSE was used to compare the performance of the developing surrogate models.} 
    \label{table:MSE_perf}
\end{table*}

To develop the NN model, all the inputs and outputs were normalized to a mean of 0 and a standard deviation of 1. This is recommended when NN is used ~\cite{murdoch2019interpretable}. In addition, the complexity of the NN model was increased until no further improvement in performance was observed. The final NN model consisted of 4 hidden layers, REUL activation, and Adam optimizer. Ketker \cite{ketkar2017introduction} recommends using RELU activation and Adam optimizer in similar applications. Furthermore, to improve the performance of the model, the maximum and minimum melt pool geometries were scaled, as well as a  classification step was introduced that splits the results into four classes:

\begin{enumerate}
    \item Samples that don’t melt. The maximum temperature is lower than the melting temperature. 
    \item Melting occurs in the length and width dimensions. There is no melting in the depth dimension as the maximum temperature is close to the melting temperature. 
    \item Melting occurs in 3 dimensions. The melt pool depth has a maximum of 9 $\mu$m. 
    \item Melting occurs in 3 dimensions. The melt pool depth is larger than 9 $\mu$m.
\end{enumerate}

The classification steps were introduced to improve the performance of the model. Furthermore, to account for the parameter uncertainty, model discrepancy, and measurement error, which are expressed in the canonical Kennedy and O'Hagan (KOH) framework ~\cite{kennedy2001bayesian}:

\begin{equation}\label{eq:KOH}
    y^{E} = \tau y^{S} (\textbf{x},\theta^{*}) + \delta(\text\bf{x}) + \epsilon
\end{equation}

Where $\tau$ is a scaling parameter, y$^{E}$ is the experimental observation, y$^{S}$ is the prediction of the surrogate model, $\delta$ is the discrepancy function, and $\epsilon$ is the measurement error which follows an independently distributed Gaussian noise with a mean of 0. Furthermore, \textbf{x} is the control input, and $\theta^{*}$ is the estimate of the fixed by unknown calibration parameters $theta$. In Bayesian modeling, the estimates of $\theta$ and model hyperparameters can be derived from the posterior distribution. 

However, since the NN surrogate model is a deterministic ML model, we assume that $\tau$ = 1 and $\delta$ follows a Gaussian process with a mean of 0 and a co-variance c($\cdot$,$\cdot$), therefore Equation \ref{eq:KOH} can be rewritten as the following:

\begin{equation}\label{eq:KOH_NN}
    y_{E}|\textbf{x},\theta^{*},\phi \sim \mathcal{GP}(y^{S}(\textbf{x},\theta^{*}),c(\cdot,\cdot)) + \mathcal{N}(\textbf{0},\sigma^{2}_{\epsilon}\mathcal{I})
\end{equation}

The calibration parameter $\theta$ and the hyperparameters $\phi$ are estimated separately to calibrate the NN surrogate model. To estimate $\theta$, $\theta^{*}$ is expressed as the optimal value and minimizes the sum of squared residuals. To generate samples for $\phi$, Markov Chain Monte Carlo (MCMC) is used and $\phi^{*}$ can be defined as the maximum posterior probability (MAP) estimate.With $\theta^{*}$ and $\phi^{*}$, the melt pool dimensions can be predicted. 

\begin{figure}
    \centering
    \begin{subfigure}[b]{.45\linewidth}
        \includegraphics[width=\linewidth]{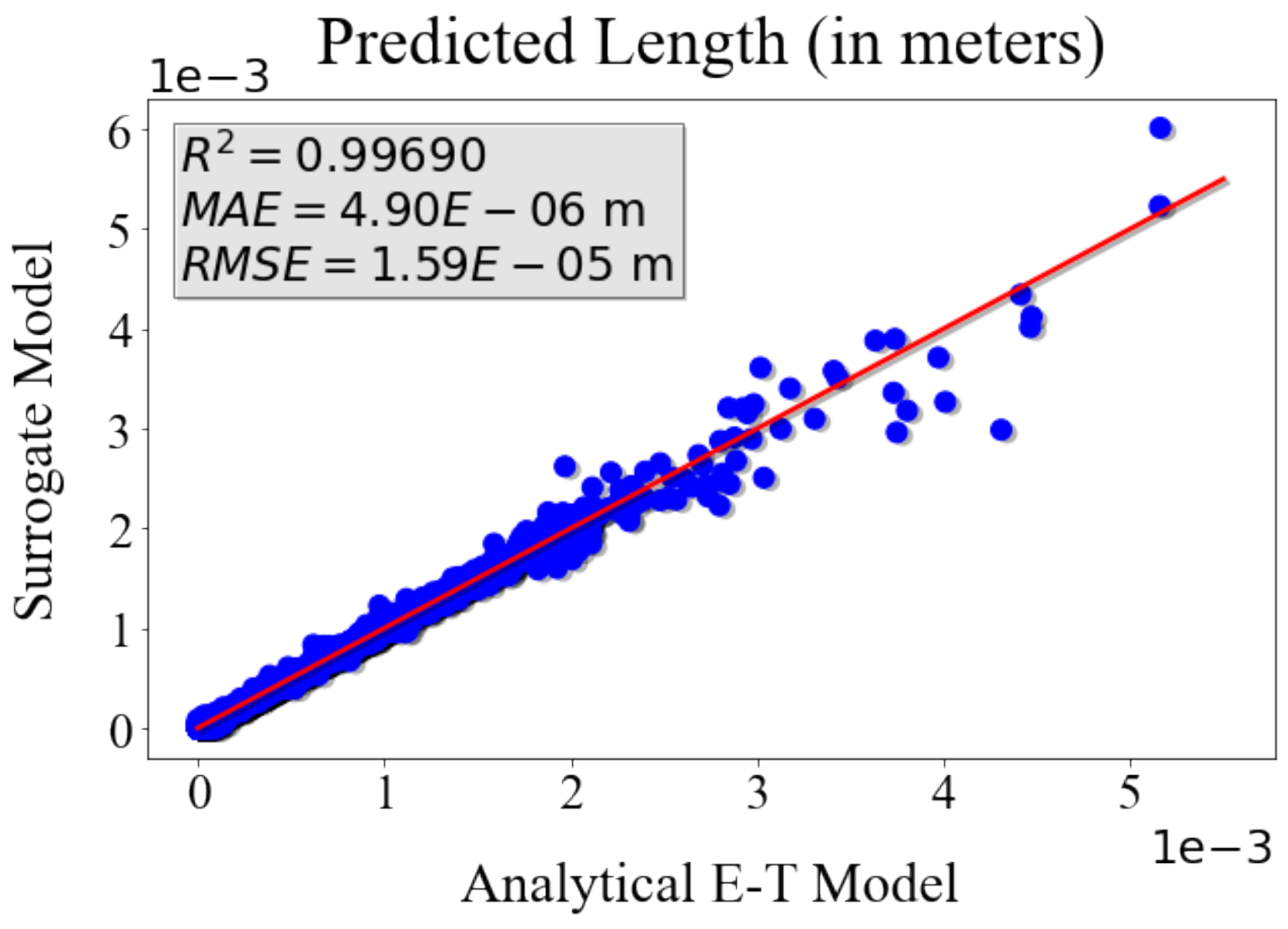}
        \setcounter{subfigure}{0}%
        \caption{Predicted melt pool length}\label{fig:pred_l}
    \end{subfigure}
    \begin{subfigure}[b]{.45\linewidth}
        \includegraphics[width=\linewidth]{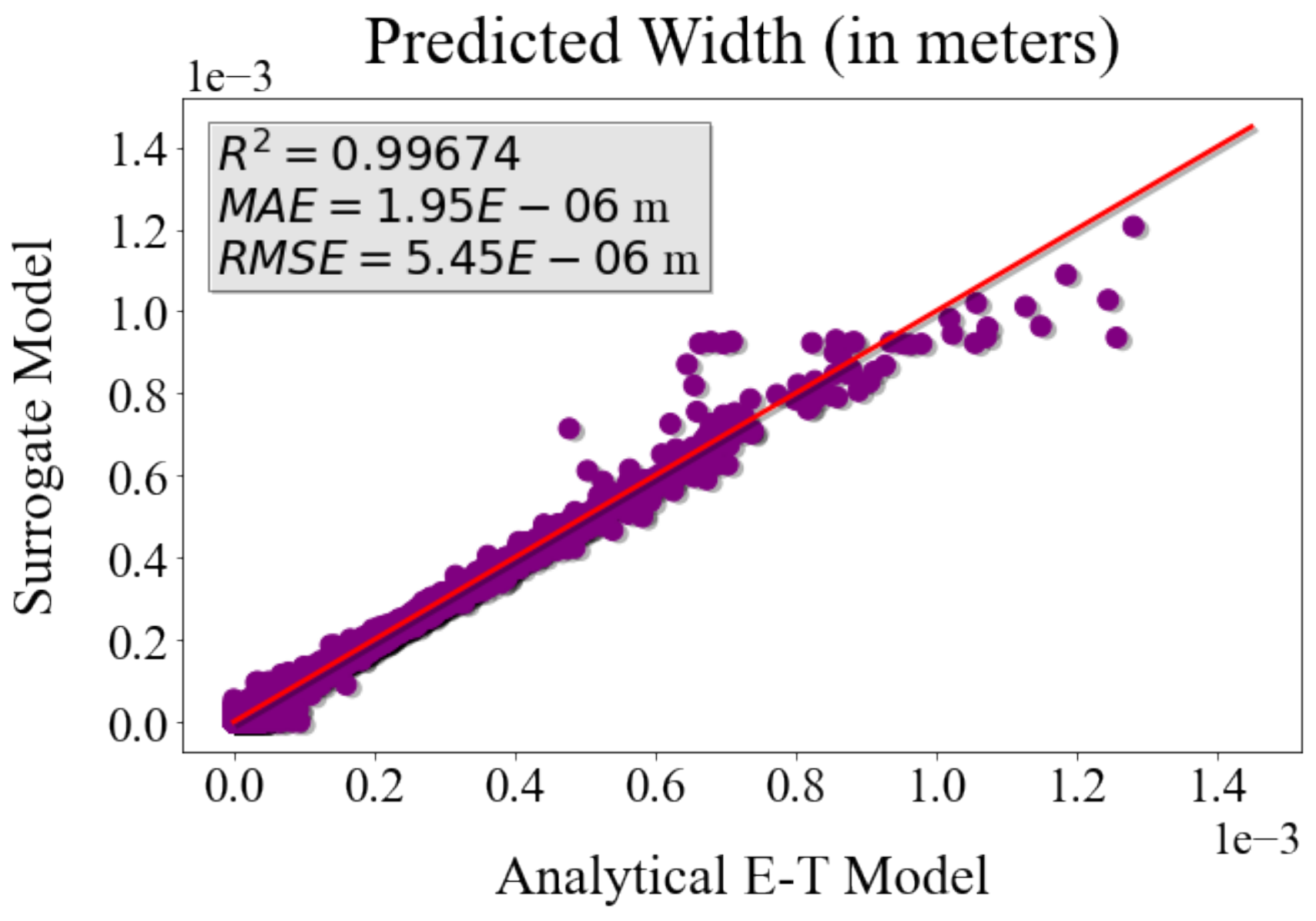}
        \setcounter{subfigure}{2}%
        \caption{Predicted melt pool width}\label{fig:pred_w}
    \end{subfigure}
    \begin{subfigure}[b]{.45\linewidth}
        \includegraphics[width=\linewidth]{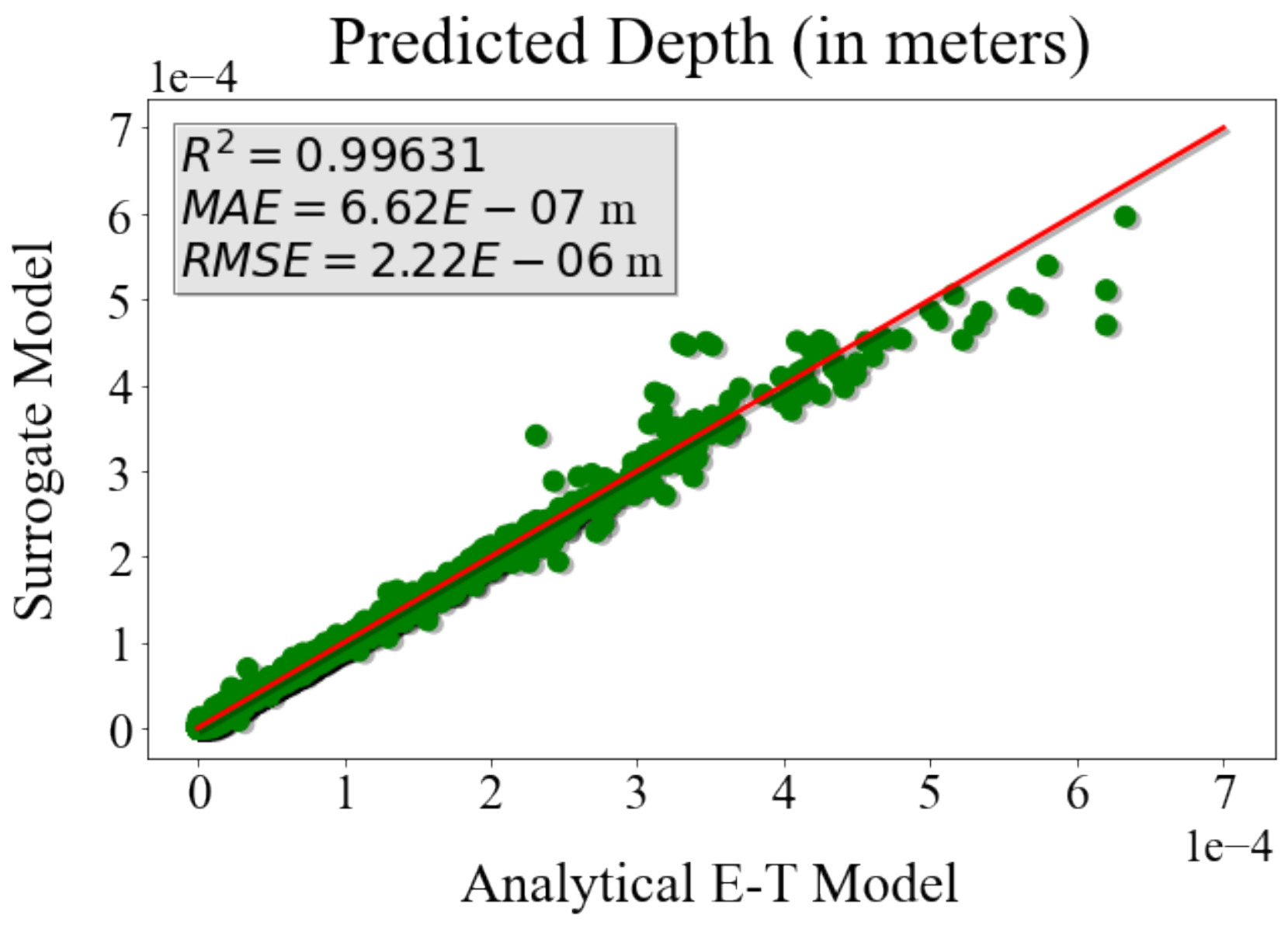}
        \setcounter{subfigure}{2}%
        \caption{Predicted melt pool depth}\label{fig:pred_d}
    \end{subfigure}
    \begin{subfigure}[b]{.45\linewidth}
        \includegraphics[width=\linewidth]{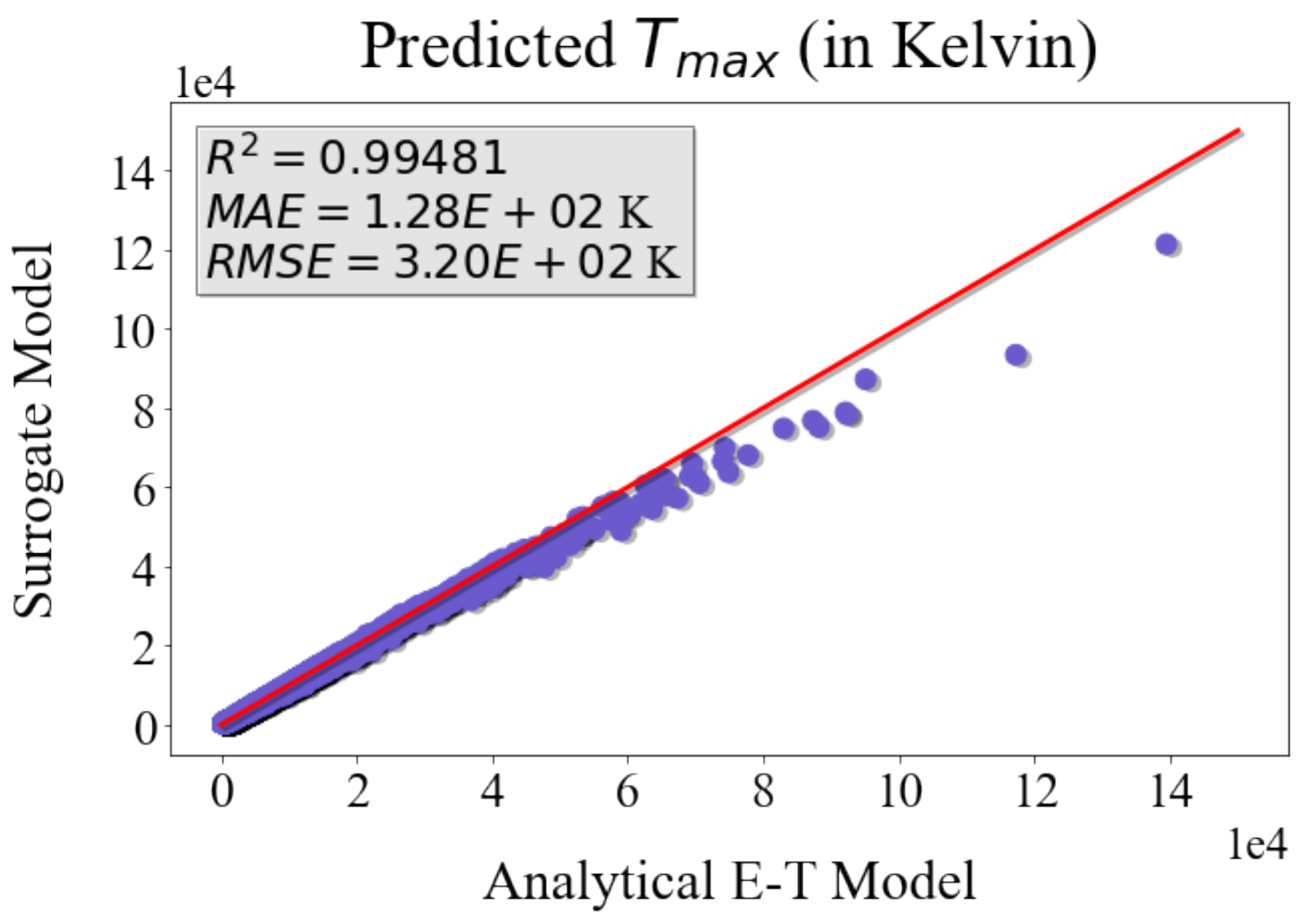}
        \setcounter{subfigure}{2}%
        \caption{Predicted melt pool maximum temperature}\label{fig:pred_Tmax}
    \end{subfigure}
\caption{The NN surrogate model predictions are compared to the analytical E-T model predictions. Using the performance metrics R$^{2}$, MAE, and RMSE, it can be analyzed that the NN model can accurately predict the melt pool dimensions at a faster computational time.}
\label{Neural Network Model}
\end{figure}

The NN surrogate model was able to perform 550,000 simulations in 120 seconds compared to 45 days using the analytical E-T model. The results of the surrogate model in predicting the melt pool geometry and maximum temperature are shown in Figure \ref{Neural Network Model}. Three different performance metrics were used to evaluate the performance of the NN surrogate model in predicting the melt pool dimensions: root mean squared error (RMSE as defined in Equation \ref{eq:RMSE} where y$_{i}^{E}$ is the expected prediction from the analytical E-T model), mean absolute error (MAE as defined in Equation \ref{eq:MAE}) and coefficient of determination (R$^{2}$ as defined in Equation \ref{eq:R2}). The parameters, RMSE and MAE, are used to evaluate the predictions as the length, width and depth can vary from units to hundreds over the parameter and/or composition space. In addition R$^{2}$ is also used to show how well the NN surrogate model can predict the melt pool geometries compared to the predictions of the analytical E-T model. From Figure \ref{Neural Network Model}, it can be seen that the R$^{2}$ value is around 0.99 for melt pool length, width, and depth, while the MAE and RMSE are in the order of 10$^{-5}$-10$^{-7}$ meters for the geometry. 

\begin{equation}\label{eq:RMSE}
    RMSE = \sqrt{\frac{\sum_{i=1}^{N_{P}} (y_{i}^{E} - y_{i}^{P})^{2}}{N_{P}}}
\end{equation}

\begin{equation}\label{eq:MAE}
    MAE = \frac{\sum_{i=1}^{N_{P}} (y_{i}^{E} - y_{i}^{P})^{2}}{N_{P}}
\end{equation}

\begin{equation}\label{eq:R2}
    R^{2} = 1 - \frac{\sum_{i=1}^{N_{P}} (y_{i}^{E} - y_{i}^{P})^{2}}{\sum_{i=1}^{N_{P}} (y_{i}^{E} - y_{i}^{Ave})^{2}}
\end{equation}

Therefore, the calibrated NN surrogate model has a good performance, making it a faster method to estimate the melt pool dimensions for L-PBF AM.

\subsection{G-S Model for the Keyhole Depth}
Among the defect modes in AM, keyholing occurs at high laser power and low scanning speeds; i.e: too high input energy density~\cite{SCIPIONIBERTOLI2017331}.
It is characterized by the formation of pores due to the entrapment of shielding gas and a melt pool with a high depth-to-width ratio.

Therefore, analytical thermal models, such as E-T model, used in calculating the melt pool dimensions fail when keyholing occurs since they are formulated in such a way as to estimate the heat conduction mode only. For this reason, the simple expression proposed by Gladush and Smurov~\cite{gladush2011physics} (referred to as G-S model) is applied in this work to estimate the keyhole depth, as follows:

\begin{equation}
\label{eq: G-S model}
D_k = \frac{AP}{2 \pi \kappa T_{b}} ln\bigg(\frac{d+\frac{\alpha}{\nu}}{d}\bigg)
\end{equation}

where $A$ is the laser absorptivity coefficient, $P$ is the laser power [W], $\kappa$ is the thermal conductivity at [W/m.K], $T_{b}$ is the boiling temperature [K], $d$ is the laser beam size [m], $\nu$ is the scanning speed [m/s].

\subsection{Defect Description and Criteria}
Various criteria for lack-of-fusion, balling, and keyholing proposed in the literature are evaluated to define regions in the \emph{process spaces} where the defects may occur. In addition to criteria that describe defects in processing parameter spaces, there also exist indicators that are solely functions of \emph{alloy chemistry}. Both categories of criteria are discussed further in this section.

\subsection{Processing-Informed Lack-of-Fusion Criterion}
For lack-of-fusion, two sets of criteria were leveraged (Equation \ref{lof1} and Equation \ref{lof2}), where $t$ is the powder layer thickness. Lack-of-fusion pores occur when insufficient energy is deposited onto the material, which results in incomplete bonding with the underlying layers. Therefore, if the melt pool depth is less than the powder layer thickness, the melt pool will not fully be bonded to the substrate and/or previous layer leading to lack-of-fusion pores. Equation \ref{lof1} captures this geometric constraint~\cite{zhang2021efficient}. Equation \ref{lof2} defines the second criterion for lack-of-fusion introduced by Zhu et al.~\cite{zhu2021predictive}. In this case, it is stated that lack-of-fusion may occur when the hatch spacing, $h$, is greater than the maximum hatch spacing, $h_{max}$~\cite{zhu2021predictive,SEEDE2020199} beyond which porosity due to insufficient overlapping melt tracks occurs. To avoid lack-of-fusion-induced porosity, the hatch spacing needs to provide a good joint between adjacent tracks to avoid incomplete bonding. The maximum hatch spacing can be calculated using Equation \ref{hmax}~\cite{SEEDE2020199}. Using Equation \ref{hmax}, it is possible to estimate the maximum hatch spacing that would still result in fully dense layers at different locations of the process map.

					\begin{equation}\label{lof1}
					    D \leq t
					\end{equation}
					
					\begin{equation}\label{lof2}
					    (\frac{h}{W})^{2} + \frac{th}{th+D} \geq 1
					\end{equation}
					
						\begin{equation}\label{hmax}
					    h_{max}  = W\sqrt{1-\frac{t}{t+D}}
					\end{equation}

\subsection{Processing-Informed Balling Criteria}
Balling occurs when the melt pool breaks into droplets instead of a continuous pool due to Plateau-–-Rayleigh capillary instability. The instability is observed at high laser power and scan speeds and can be evaluated using a criterion based on the ratio of the melt pool length and width ($L/W$), as defined in Equation \ref{ball1}. The second criterion implemented is that proposed by Yadroisev et al.~\cite{YADROITSEV20101624}. For single tracks, they showed that the threshold value between the stability zone and the instability zones of a melt pool can be defined using Equation \ref{ball2}.

\begin{equation}\label{ball1}
    \frac{L}{W} \geq 2.3
\end{equation}

\begin{equation}\label{ball2}
    \frac{\pi W}{L} < \sqrt{\frac{2}{3}}
\end{equation}

\subsection{Processing Informed Keyhole Criteria}

At combinations of high power and low velocity, keyhole-induced pores can form due to rapid evaporation of the molten liquid that causes a deep penetration of the molten material by the recoil pressure induced on the melt pool surface, leading to the formation of vapor cavities. Consequently, the resultant melt pool is larger than the depth formed during the conduction mode. An experimentally-obtained criterion has been suggested in terms of the melt pool geometry~\cite{JOHNSON2019199,SEEDE2020199}, defined in Equation \ref{key1}.

However, King et al.~\cite{king2014observation} introduced another criterion for keyholing that considers not only the geometry of the melt pool but also the material properties in addition to processing parameters. They showed a positive correlation between the normalized enthalpy and the melt pool depth which in turn is correlated to keyholing. As shown in Equation \ref{key2}, the criterion utilizes the specific enthalpy, $h_s = \rho C_p T_{liquidus}$ ($\rho$ is the density [kg/m$^3$], $C_p$ is the effective specific heat [J/kg K] and $T_{liquidus}$ is the liquidus temperature). In the original formulation by King et al.~\cite{king2014observation}, the heat capacity for the solid phase was considered. However, the heat capacity for the solid phase underestimates the energy absorbed by the metal from the laser. Therefore, we replaced the heat capacity with the effective heat capacity to account for both the sensible and latent heat of the given alloy from room temperature until the liquidus temperature. 

Recently, Gan et al.~\cite{gan2021universal} introduced a 'universal' keyholing criterion that can define regions of conduction, transition, and keyholing mode in the melt pool. This dimensionless keyholing criterion, $Ke$, as defined in Equation \ref{Ke}, was derived using dimensional analysis and by modifying the Buckingham-Pi theorem. The criterion $Ke$ utilizes processing parameters such as the laser power ($P$ [W]), velocity ($\nu$ [m/s]), and beam radius ($r_0$ [m]) as well as material properties, i.e., the absorptivity ($\eta$), liquidus temperature ($T_{liquidus}$ [K]), substrate temperature ($T_0$ [K]), density ($\rho$ [kg/m$^3$]), heat capacity ($C_p$ [J/kgK]), and thermal diffusivity ($\alpha$ [m$^2$/s]). In their work, Gan et al.~\cite{gan2021universal} used experiments to derive a threshold value of $Ke > 6.0$ for the keyhole mode in the L-PBF processes. 

\begin{equation}\label{key1}
\frac{W}{D} \leq 2.5
\end{equation}

\begin{equation}\label{key2}
   \frac{\Delta H}{h_s} = \frac{AP}{\pi h_s \sqrt{\alpha va^3}} > \frac{\pi T_{boiling}}{T_{liquidus}}
\end{equation}

\begin{equation}\label{Ke}
    Ke = \frac{\eta P}{(T_{liquidus} - T_0) \pi \rho C_p \sqrt{\alpha \nu r^3_{0}}}>6
\end{equation}

\subsection{Summary of Processing-Informed Criteria}
In our framework, two criteria for lack-of-fusion, two for balling, and three for keyholing, for a total of twelve different combinations of these defect criteria are considered. Table \ref{table:labels} summarizes the different criteria for lack-of-fusion, balling, and keyholing. When constructing a printability map, in addition to the defect criteria, contour lines that define the maximum hatch spacing allowed to prevent porosity due to a lack of overlap of the melt pool tracks (defined in Equation \ref{hmax}) are plotted on the printability maps as well. 

\renewcommand{\arraystretch}{1.5}
\begin{table*}
    \centering
    \setlength\extrarowheight{5pt}
    \begin{tabularx}{0.9\textwidth} { 
       >{\centering\arraybackslash}X 
       >{\centering\arraybackslash}X 
       >{\centering\arraybackslash}X  }
     \hline
     \textbf{Defect Type} & \textbf{Label} & \textbf{Equation} \\
    \hline
    Lack-of-Fusion & LOF1 &     D $\leq$ t \\
 
    & LOF2 &  ($\frac{h}{W})^{2}$ + $\frac{th}{th+D} \geq$ 1 \\
    \hline
    Balling & Ball1 &  $\frac{L}{W}$ $\geq$ 2.3 \\

    & Ball2 & $\frac{\pi W}{L} < \sqrt{\frac{2}{3}}$ \\
    \hline
    Keyholing  & KH1  & $\frac{W}{D}$ $\leq$ 2.5  \\

    & KH2 & $\frac{\Delta H}{h_s} = \frac{AP}{\pi h_s \sqrt{\alpha va^3}} > \frac{\pi T_{boiling}}{T_{liquidus}}$ \\

    & KH3 & $Ke = \frac{\eta P}{(T_{liquidus} - T_0) \pi \rho C_p \sqrt{\alpha \nu r^3_{0}}}>6$ \\
    \hline
    \end{tabularx}
    \caption{The equation and labels for each respective defect used to graph the printability maps in Figure \ref{fig:Ana_ET}, Figure \ref{fig:NN_ET}, and Figure \ref{fig:Scaled_ET}}. 
    \label{table:labels}
\end{table*}

\subsection{Composition-Informed Hot Cracking Susceptibility}

Hot cracking, also known as hot tearing, is the formation of fissures in casting due to solidification shrinkage and thermal strains arising during solidification and subsequent cooling~\cite{yan2006prediction}. Hot cracking is mitigated by sufficient liquid feeding to the mushy zone and a high grain growth rate. Such hot cracking typically occurs during the late stages of solidification, where liquid feeding is restricted, and grain growth slows down~\cite{KOU2015366,yan2006prediction}. Several indicators of hot cracking exist and have been used in alloy design \cite{BAE2014361,FARKOOSH2013596}. One common hot cracking indicator is the solidification range, defined as the difference between an alloy's solidus and liquidus temperatures. Wide solidification ranges have been shown to increase the tendency of hot cracking \cite{WANG201390}. Wide solidification ranges have also been shown to increase the tendency of compositional microsegregation during alloy solidification \cite{wangInvestigation2009}.

Furthermore, several hot cracking criteria exist based on the Scheil-Gulliver curve. The Scheil-Gulliber curve describes how the liquid phase fraction varies with temperature ($f_L-T$) during non-ideal solidification. One such criterion was proposed by Clyne and Davies~\cite{yan2006prediction}. This hot cracking susceptibility coefficient is the ratio between the time the casting is vulnerable to cracking, $t_v$, and the time the casting can relieve thermal stresses, $t_R$~\cite{clyne1981influence,yan2006prediction}, as shown in Equation~\ref{t_v vs. t_R}. As mentioned, hot cracking typically occurs during the late stages of solidification when liquid feeding is restricted. A Scheil-solidification simulation is applied to generate heat-evolution vs. temperature and liquid fraction vs. temperature curves to estimate these critical timescales. By assuming that $dQ/dt \propto 1/\sqrt{t}$, a liquid fraction vs. time curve ($f_L-t$) is obtained. This $f_L-t$ curve determines the critical timescales $t_v$ and $t_R$. Following the convention established by Clyne and Davies, $t_R$ is the period that $f_L$ is between 0.6 and 0.1, whereas $t_v$ is the period that $f_L$ is between 0.1 and 0.01~\cite{clyne1981influence}. It should be noted that larger $t_v$/$t_R$ corresponds to an increased frequency of hot cracking~\cite{yan2006prediction,clyne1981influence}.

\begin{equation}
\label{t_v vs. t_R}
  CSC = \frac{t_{v}}{t_{R}}
\end{equation}

\subsection{Chemistry-Informed Balling Susceptibility}

Balling is an AM defect mode where an elongated melt pool suffers from instability and segregates into disconnected beads or balls \cite{ZHOU201533,DEBROY2018112}. Balling has been shown to be influenced by both processing parameters and material properties. As such, this phenomenon can be described by simplified models based on melt pool geometry \cite{YADROITSEV20101624} or the intrinsic materials properties of the printed material \cite{ZHOU201533}.

Regarding materials properties, Zhou et al. predicted balling in tungsten using a simple model where melt spreading and solidification compete \cite{ZHOU201533}. When the characteristic spreading time ($\tau_{spread}$) is much longer than the characteristic solidification time ($\tau_{solid}$), melt spreading is likely to be arrested by solidification before the melt pool has successfully spread and wet the underlying surface, resulting in balling \cite{ZHOU201533}. Both of these timescales are functions of intrinsic material properties. In a previous work \cite{vela2022evaluating} we further developed this notion, utilizing $\tau_{solid} / \tau_{spread}$
as a physics-based indicator capable of predicting balling in an L-PBF specimen. The predictive ability of this indicator was validated against an AM database consisting of 2,527 unique alloy-process observations and 158 unique compositions. A larger $\tau_{solid} / \tau_{spread}$ was shown to be correlated with a smaller balling region in the printability map associated with an alloy.

This model is based on several assumptions. The spreading and solidification of the droplet are considered separately. Assuming conduction is the dominant heat transfer mode, the characteristic solidification time consists of two distinct regimes, $\tau_{1}$ and $\tau_{2}$. In the first regime (Equation \ref{eq:tau1}),  $\tau_{1}$ is the time required to remove heat from a spherical droplet with radius $a$ via conduction to the underlying substrate. The second regime (Equation \ref{solid_tau2}) describes the time to remove latent heat of fusion $L$ and solidify. The total solidification time is given by Equation \ref{eq:tau_solid}, where the factor of 2 accounts for the fact that approximately half of the droplet surface is in contact with the underlying substrate, halving the heat transfer area and approximately doubling the solidification time. Assuming isothermal and homologous wetting of a molten droplet on a flat surface where both solidification and oxidation are neglected, spreading is primarily driven by capillary flow. The characteristic spreading time is then estimated from Equation \ref{eq:tau_spread} \cite{ZHOU201533}. The initial radius of the droplet, $a$, is assumed to be 100 $\mu$m, common during L-PBF \cite{ZHOU201533,vela2022evaluating}. The fusion temperature, $T_f$ is assumed to be distinct and is taken as the material's solidus temperature \cite{vela2022evaluating}. The ambient temperature, $T_a$ is taken as $25^\circ$C.

\begin{equation} \label{eq:tau1}
  \tau_1 = \frac{a^2k}{3\alpha k_a}\ln\left(\frac{T_o-T_a}{T_f-T_a}\right)
\end{equation}

\begin{equation}
  \tau_2 = \frac{a^2k}{3\alpha k_a}\left(1+ \frac{k_a}{2k}\right)\frac{L}{C(T_f - T_a)}
  \label{solid_tau2}
\end{equation}

\begin{equation} \label{eq:tau_solid}
  \tau_{solid} = 2(\tau_1 + \tau_2)
\end{equation}

\begin{equation}  \label{eq:tau_spread}
  \tau_{spread} = \left(\frac{\rho a^3}{\sigma}\right)^{1/2}
\end{equation}

\subsection{Construction and Ranking of the Printability Maps}
After obtaining the melt pool profiles from one of the three thermal models previously mentioned, the criteria listed in Table \ref{table:labels} are used to determine regions in the process space associated with a lack-of-fusion, balling, and keyholing. These defect regions are defined with different combinations of criteria. These defined defect regions are used to construct printability maps. The criteria sets used to construct these maps are benchmarked by comparing the predictions in the printability maps with the available data. 

To rank the printability maps and to evaluate the various criteria combinations, we assumed that the combination of models and criteria constituted a binary classifier separating a specific defect region from the rest of the processing space---.
e.g., lack-of-fusion or not lack-of-fusion. The experimental data points corresponding to a particular label were considered the positive class, while everything else was considered the negative class. To rank the performance, a wide variety of performance metrics were used, such as precision (Equation \ref{eq:precisioin} ), recall (Equation \ref{eq:recall}), and accuracy (Equation \ref{eq:accuracy}) where P and N are points in the positive and negative class. Furthermore, TP stands for true positive or the number of positive classes that were correctly classified, TN stands for true negative or the number of negative points that were correctly classified, FP stands for false positive or the number of negative points that were misclassified, and FN stands for false negative or the number of positive points that were misclassified. 

In the context of the printability map, precision measures how accurately the prediction was made, where 100\% precision means that every point predicted to belong to a defect class truly belongs to that defect class. On the other hand, an undefined precision value corresponds to all positive points being misclassified and all negative points being correctly classified (TP and FP are 0). Recall measures the ability of the criteria combination to capture the entire region in the processing space where a specific defect occurs. A 100\% recall indicates the criteria combination was able to fully map the defect region. Finally, accuracy measures how effectively the printability maps made correct predictions for all the defect modes. 

For each printability map, the above metrics were obtained for each of the four binary classification cases and averaged to find the average performance of each set of criteria. The averaging was carried out by considering the prevalence of each type of defect within the experimental data set obtained. From now on, the weighted performance metric will be referred to as is (i.e, average accuracy will be referred to as accuracy) as it will be assumed that each metric that is being discussed has been averaged to fully assess the performance of the printability map. Accordingly, for each constructed printability map, the experimental points were overlaid to help visualize and calculate the different performance metrics as mentioned above.

\begin{equation}
    Precision = \frac{TP}{TP+FP}
    \label{eq:precisioin}
\end{equation}

\begin{equation}
    Recall = \frac{TP}{TP+FN}
    \label{eq:recall}
\end{equation}

\begin{equation}
    Accuracy = \frac{TP+TN}{TP+FP+TN+FN}
    \label{eq:accuracy}
\end{equation}

\section{Results and Discussion}

\subsection{Printability of an Arbritary Co-Cr-Fe-Mn-Ni High Entropy Alloy System}
\label{sec:Arb_cantor}

\begin{figure}[htb!]
    \centering
    \includegraphics[width = \linewidth]{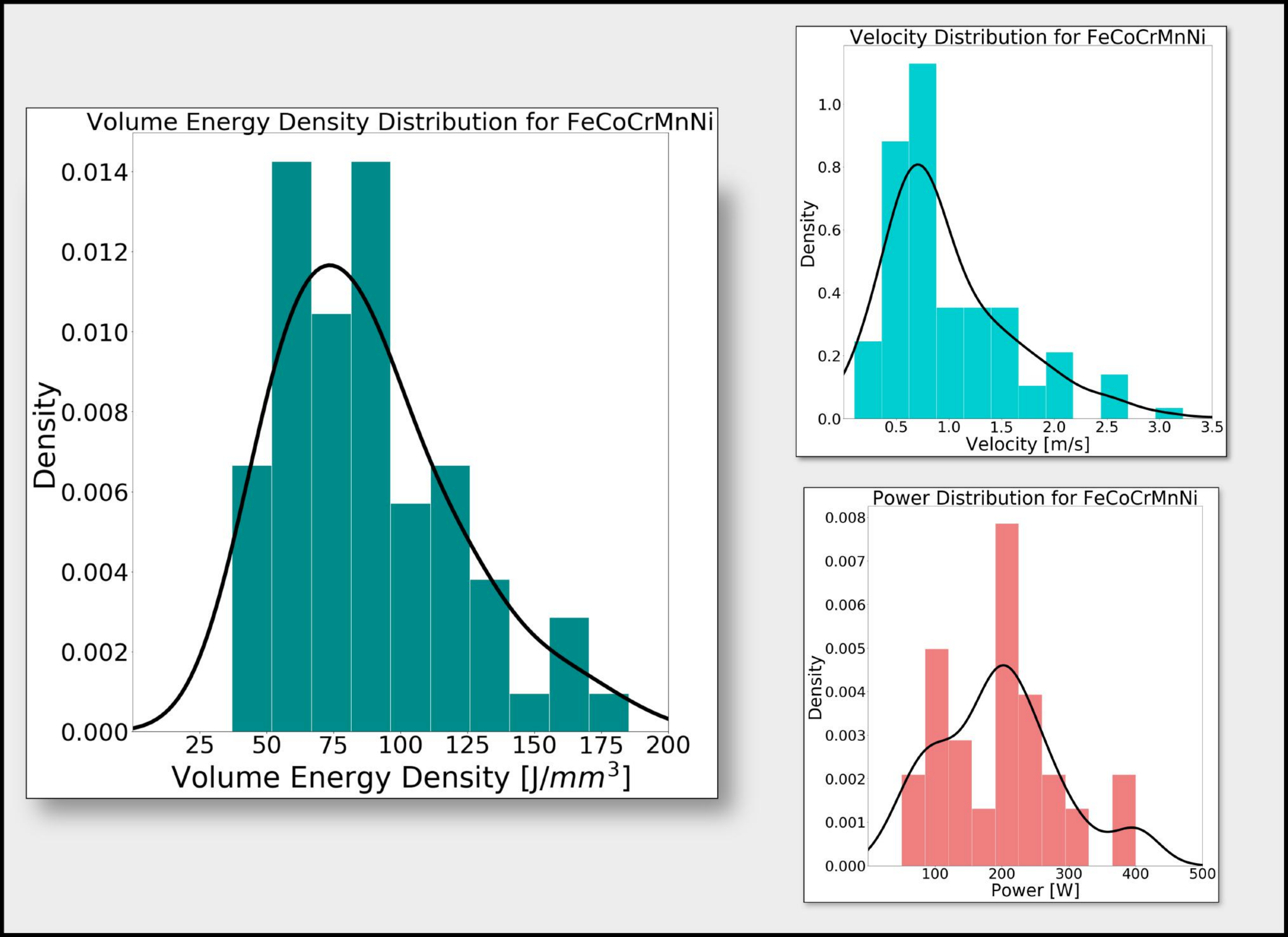} 
    \caption{The histogram and kernel density estimate (KDE) plots for power, velocity, and volume energy density (VED) show the distribution of values used to print for the equiatomic CoCrFeMnNi alloy, as reported in the literature. The density represents the probability density function of the parameter on the x-axis in the plot.}
    \label{fig:Database_Dist}
\end{figure}

To demonstrate our framework and identify a ground truth to benchmark the proposed framework, we chose the equiatomic CoCrFeMnNi alloy system for our case study, as it is one of the most thoroughly studied HEAs as shown in Figure \ref{fig:HEAs_Data_Sh}. By conducting a comprehensive data mining of the literature, we built an in-house database that captures data points from the literature in combination with experiments conducted in-house. Each entry in the database captures the chemistry-processing space with over 500 features that are related to different aspects of metal L-PBF AM that affect the build. The features range from processing conditions, material properties, and powder properties to the printing outcome, post-performance of the printed part, and more. The combination of processing parameters for the CoCrFeMnNi system that is captured within the database is shown in Figure \ref{fig:Database_Dist}. Based on the data, we chose the range of processing parameters where the power was set from 400 W to 360W, and velocity was set from 0.05 m/s to 2.5 m/s. Furthermore, due to the limited data available, we considered both single tracks as well as full builds to build a comprehensive data set that includes all of the mentioned defects.  

For the equiatomic CoCrFeMnNi alloy, twelve printability maps for each thermal model were constructed and ranked according to their accuracy. However, to show the full analysis of the performance of the individual printability maps, the precision and recall values are also presented.

For the thermal E-T model, a 17-by-14 design of experiment for power and velocity was used. However, for the dimensionless E-T thermal model, the processing space was reduced to ensure that the value of $p$ and $B$ were within the range of 0.1-5 and 1-20 in these calculations, respectively. Based on experimental data that was mined, the hatch spacing was held constant at 70 $\mu$m, and the value for the beam diameter was held constant at 80 $\mu$m. The powder layer thickness was also held constant at 30 $\mu$m.  

\begin{figure*}
    \centering
    \includegraphics[width=1\textwidth]{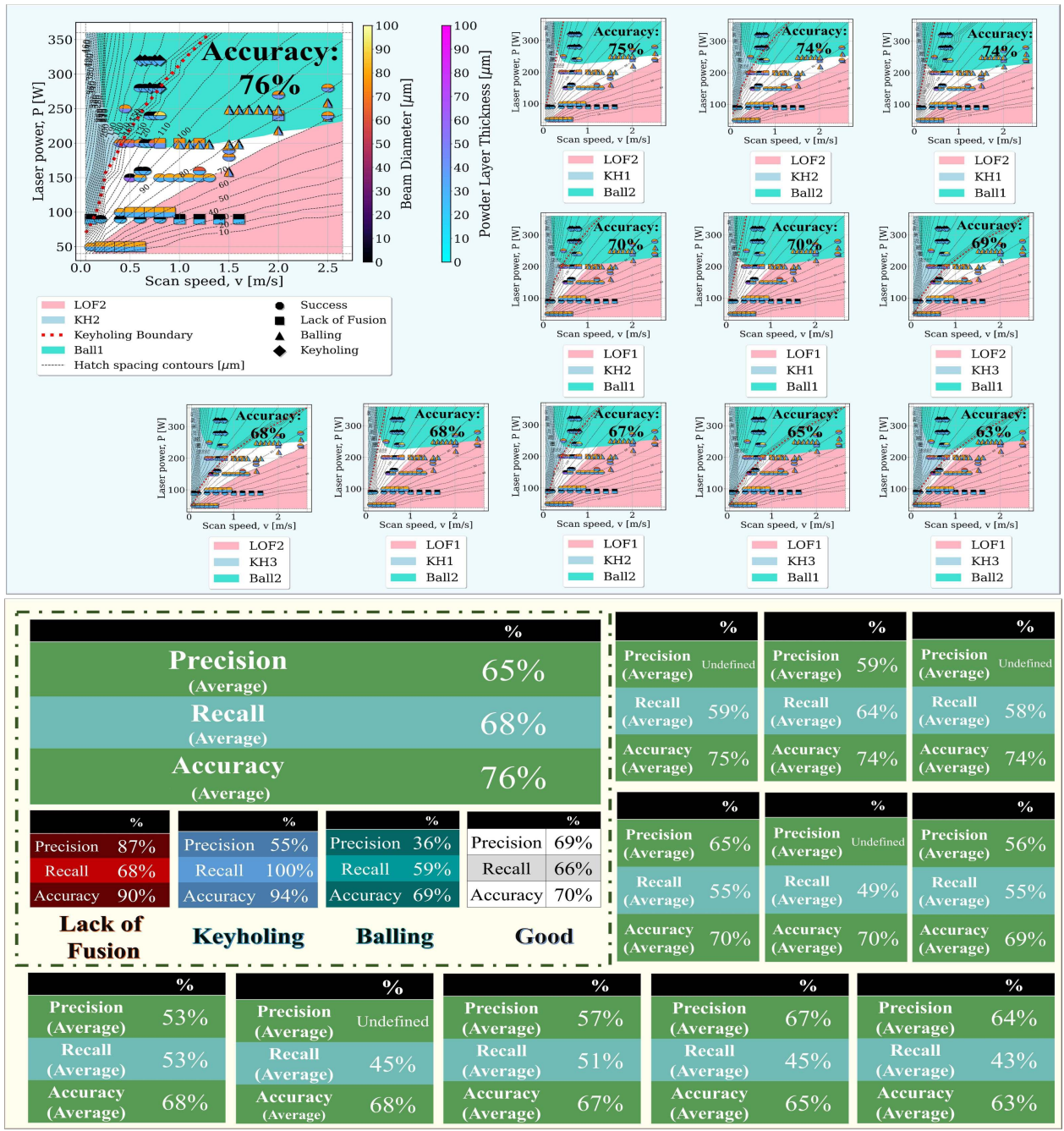}
    \caption{Using the NN E-T model, the highest accuracy that was obtained was 76\% for the set of criteria of LOF2, Key2, and Ball1. It can be noted that the results are similar, if not the same, as that obtained from the analytical E-T model. However, the NN E-T model is computationally less expensive than that of the analytical E-T thermal model. The printable index is 18.5\%.}
    \label{fig:NN_ET}
\end{figure*}

For the analytical E-T model and the NN E-T model, the highest accuracy of 76\% was obtained as shown in Figure \ref{fig:Ana_ET} and Figure \ref{fig:NN_ET}. The values for the highest accuracy were obtained with the criteria set using Equation \ref{lof2}, Equation \ref{key2}, and Equation \ref{ball1} for their respective defects. Furthermore, the relative size of  the printable region (i.e., the percentage of the map that is white) was evaluated using the Python packages OpenCv and Sci-kit Image. The area of the printable region will be referred to as the \emph{printable index}. The printability maps with the highest accuracy generated using the analytical E-T thermal model had a printable index of 17.1\%. The NN E-T thermal model with the highest accuracy had a printable index of 18.5\%. On the other hand, the dimensionless E-T model with the highest accuracy of 72\% and a printable index of 22.2\% was obtained using the following set of criteria: Equation \ref{lof1} for lack-of-fusion, Equation \ref{key1} for keyholing and Equation \ref{ball1} for balling. 

We would like to note that the experiments that were pulled from the literature were not at the same processing values of hatch spacing, beam diameter, and powder layer thickness. Even so, we still considered these points in our calculations with the parameters we have defined in our framework. In addition, the beam diameter and powder layer thickness for each experimental point is indicated in the printability maps, where 0 indicates that the value for the particular parameter for the experimental point was not provided. Furthermore, we would like to acknowledge that across the three models, precision was a poor performance metric for keyholing, while recall was a poor performance metric for balling. It can be noted for many of the misclassifications associated with balling and keyholing, the experimental points were at the boundary of the respective defect and the neighboring regions. Therefore, the prediction was sensitive to small inaccuracies during the evaluation of the criteria set. However, this did not affect the accuracy that was used to rank the printability maps.

For the analytical and NN E-T model, the model–criteria performance was similar as the highest accuracy for both was 76\%, with the combination of optimal criteria being the same. The computational expense for the analytical E-T model was around 1.5 hours, while the computational expense for the NN E-T model was only 3 to 5 seconds, representing close to a \emph{1,000-fold acceleration at no inaccuracy cost}. On the other hand, the dimensionless E-T model had a lower accuracy, which can be attributed to the fact that due to reducing the processing space that is being sampled with the dimensionless E-T model, the quantity of the experimental points used to evaluate the performance of the printability maps was reduced in comparison to that of the previous two E-T models (i.e., about 10 experimental points were removed for the evaluation of the dimensionless E-T thermal model). 

\begin{figure*}
    \centering
    \includegraphics[width=1\textwidth]{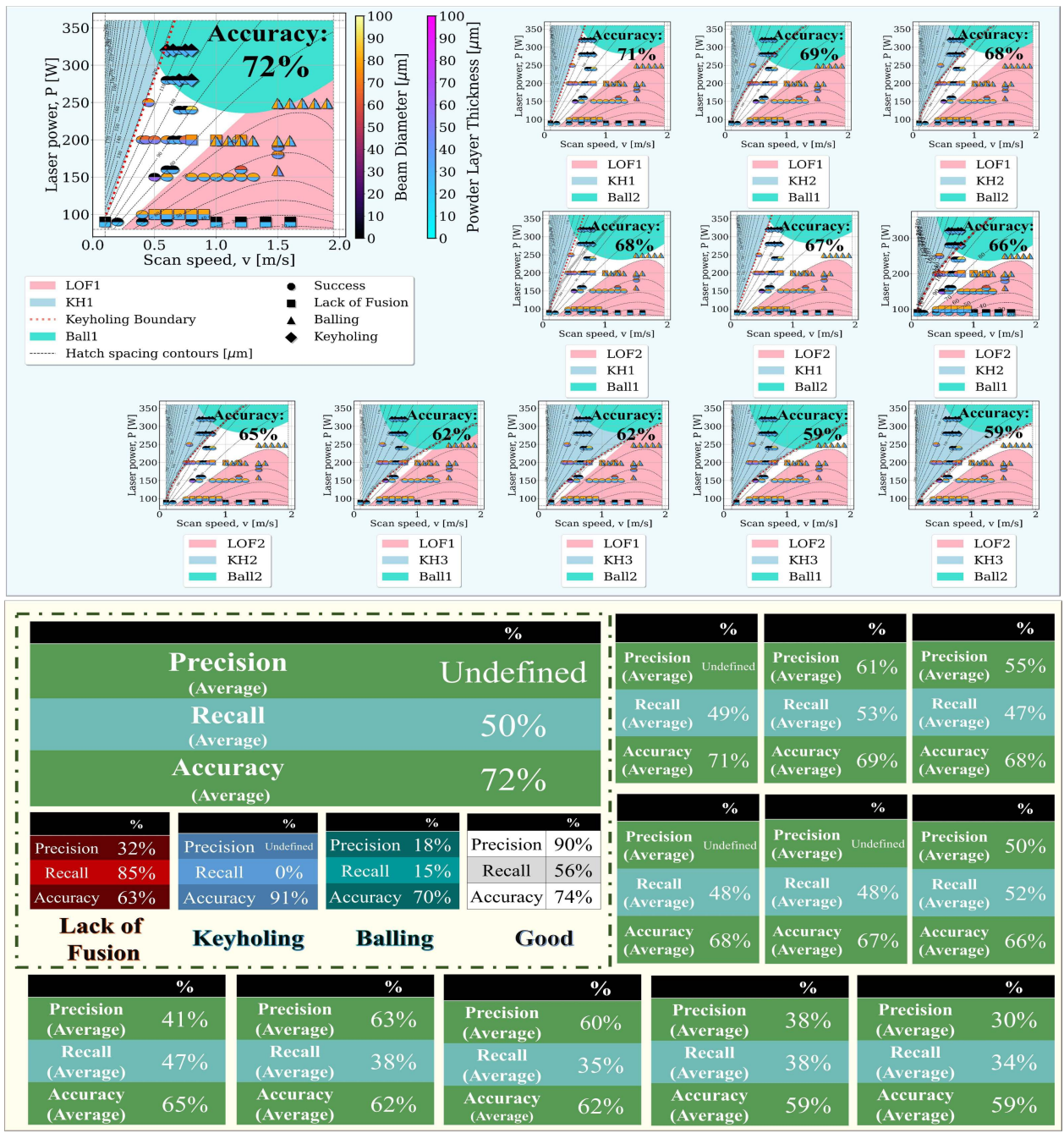}
    \caption{By obtaining the melt pool profile with the dimensionless E-T model, the highest accuracy obtained was 72\% with the optimal combination criteria of LOF1, Key1, and Ball1. The boundary for each region is a smooth boundary, which is a more realistic description of the boundary behavior than that of the analytical E-T model and NN E-T model. The printable index when using the dimensionless E-T model is 22.2\%.}
    \label{fig:Scaled_ET}
\end{figure*}

Moreover, it can be seen from Figure \ref{fig:Scaled_ET} that the boundary regions in comparison to the Figure \ref{fig:Ana_ET} and Figure \ref{fig:NN_ET} is smoother. The smoothness of the boundaries for the dimensionless E-T model is a more accurate description of reality rather than the staggering boundaries obtained from the analytical E-T and NN E-T models. Concurrently, the computational expense of the dimensionless E-T model is similar to that of the NN E-T model of 3 to 5 seconds. As stated above, the equiatomic CoCrFeMnNi alloy is the most well-researched single-phase FCC HEA. Using this alloy, we can set a 'ground truth' for HEAs that exhibit single-phase FCC phase (at elevated temperatures) structure with elements Co,Cr,Fe,Mn and Ni. Ultimately, based on our predictions, the optimal combination criteria for a general CoCrFeMnNi alloy is Equation \ref{lof2}, Equation \ref{key2}, and Equation \ref{ball1}.

\begin{figure*}
    \centering
    \includegraphics[width=1\textwidth]{Figure7.pdf}
    \caption{By obtaining the melt pool profile using the analytical E-T model, the highest accuracy of 76\% was obtained by using a hatch distance of 70 $\mu$m and beam diameter of 80 $\mu$m with criteria set of LOF2, KH2, and Ball1.The printable index is 17.1\%.}
    \label{fig:Ana_ET}
\end{figure*}

\subsection{Alloy Design for Printability and High-Temperature Applications}

While the previous analysis shows the processing window for an alloy picked at random from the vast Co-Cr-Fe-Mn-Ni alloy space, alloys can be \emph{co-designed} for high-temperature applications and amenability to AM. In order to design Cantor alloys for use in high-temperature applications, the system was screened for alloys predicted to have a single FCC phase at $1300^\circ C$ using the Thermo-Calc TCHEA5 database for high entropy alloys. In order to design printable alloys, composition-based alloy design metrics are needed. Such composition-based objectives are the hot cracking susceptibility coefficient, the $\tau_{solid} / \tau_{spread}$ balling indicator, and the solidification range $\Delta T$. 

\begin{figure}[]
    \centering
    \includegraphics[width = 1 \linewidth]{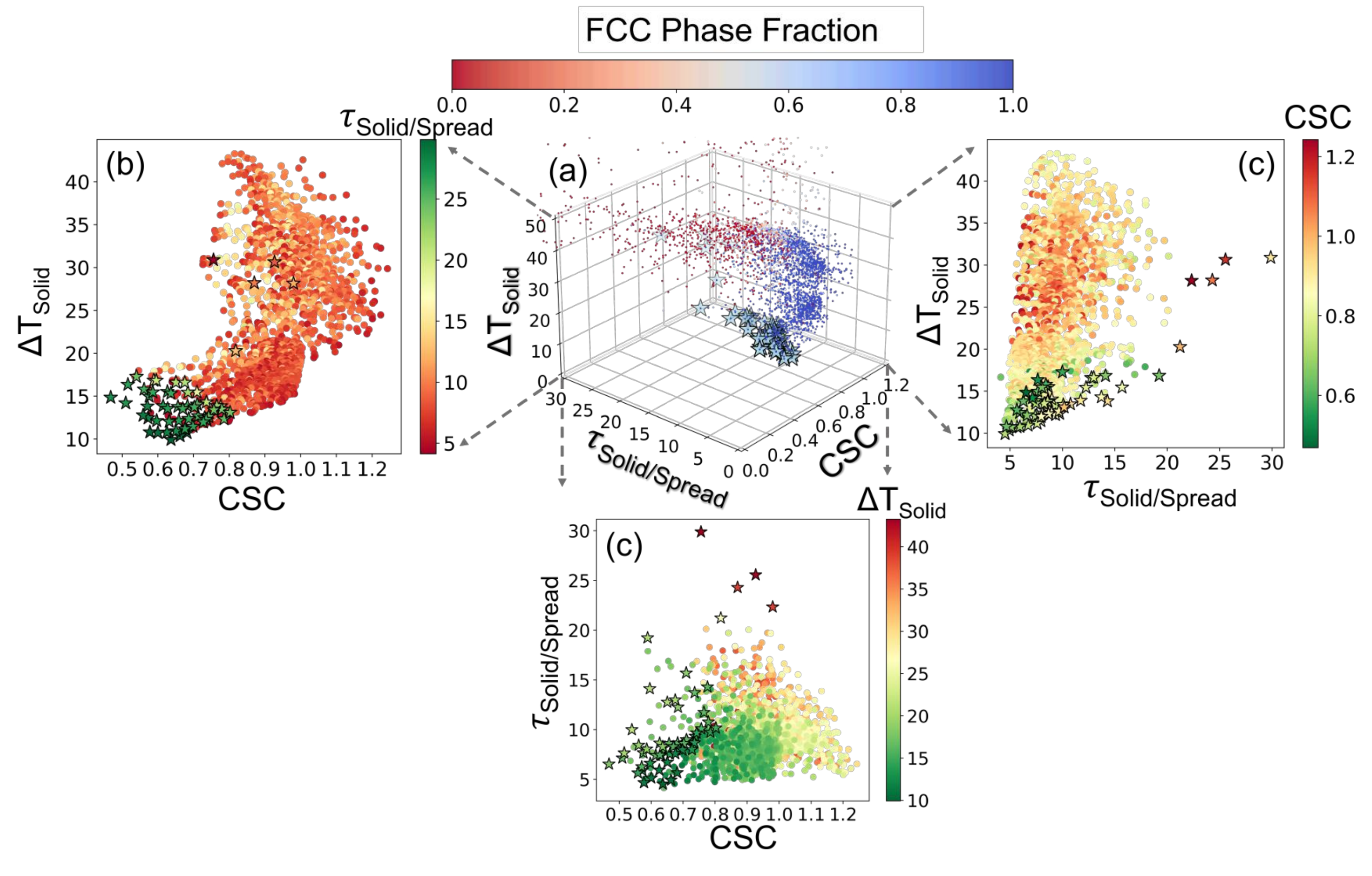}
    \caption{a) The tri-objective Pareto front is plotted in objective space. The FCC constraint is depicted as a color axis. b) The Pareto-front projected on the $\Delta T$-CSC axis with $\tau_{solid} / \tau_{spread}$ as the color axis. c) The Pareto-front projected on the $\tau_{solid} / \tau_{spread}$-CSC plane with $\Delta T$ as the color axis. d) The Pareto-front projected on the $\Delta T$-$\tau_{solid} / \tau_{spread}$ axis with CSC as the color axis.}
    \label{Pareto_3d}
\end{figure}

The aforementioned performance constraint and three printability metrics were calculated for \emph{all quinary alloys} within the Cantor system (3,876 alloys in total). 58.46\% of the space failed the HT phase stability constraint, depicted as red in  Figure \ref{Pareto_3d}a. Within the feasible space, depicted in blue, a Pareto-front of optimal alloys for AM was identified, as shown in Figure \ref{Pareto_3d}a. The 2-dimensional projections of this 3-dimensional objective space are shown in Figure \ref{Pareto_3d}b c and d. Color is used to depict the 3rd objective.

\begin{figure}[htb!]
  \centering
  \includegraphics[width=1\linewidth]{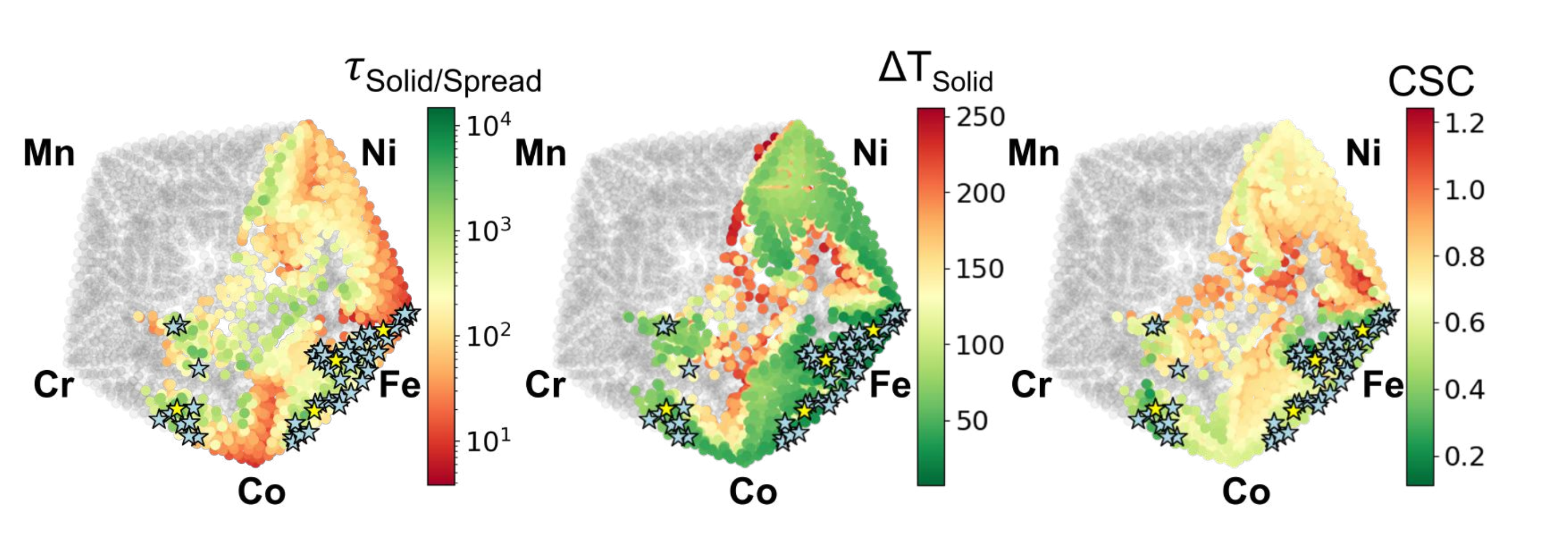}  
    \caption{The tri-objective Pareto front plotted in composition space. Alloys that fail the FCC constraint are depicted in grey. Both printability metrics, (a) the cracking coefficient and (b) $\tau_{solid} / \tau_{spread}$ plotted over t-SNE projections of the CoCrFeMnNi alloy space show how the printability objectives vary with composition.}
  \label{Pareto_tsne}
\end{figure}

In order to visualize how printability metrics vary within the composition space, we rely on a dimensional reduction technique known as t-distributed stochastic neighbor embedding (t-SNE) to project the 5-dimensional Cantor alloy space into a 2-dimensional embedding, as shown in Figure \ref{Pareto_tsne} a and b. With t-SNE, points near each other in high-dimensional space are projected near each other in the 2-dimensional embedding, however, no quantitative inferences about the relationship between points can be made by the Euclidean distance in the 2-dimensional embedding. In this work, each point in the t-SNE embedding represents an alloy with a unique composition. The points are arranged in such a way that points close to the corners of the "pentagonal" shape approach unary alloys. For example, the alloy in the Cr-corner of the t-SNE corresponds to the unary (i.e., pure) Cr; Alloys along the edge of the "pentagonal" shape that connects Cr to Mn are binary alloys consisting of Cr-Mn.

Upon inspection of Figure \ref{Pareto_tsne} regions rich in FCC-forming elements Mn and Cr do not pass the phase stability constraint and are thus depicted as gray in the tSNE projections. Regions rich in Co-Fe and Cr-Co have the optimal trade-off between balling resistance and hot cracking resistance. A k-medoids-based sampling scheme grouped the Pareto front into five representative clusters based on composition; The printability of these medoids (representative members of the cluster, depicted as yellow stars) was analyzed with the proposed framework and their printability indices were queried.

\subsection{Printability Maps for the Selected Alloys}

After analyzing the tri-objective Pareto front and using a k-medoids sampling, five alloys were analyzed with our proposed framework and the optimal combination of porosity-defect criteria that was determined from the analysis of the equiatomic CoCrFeMnNi alloy. The five alloys that were sampled include:
\begin{enumerate}
    \item Fe$_{60}$Co$_{15}$Cr$_{15}$Mn$_{5}$Ni$_{5}$
    \item Fe$_{25}$Co$_{45}$Cr$_{20}$Mn$_{5}$Ni$_{5}$
    \item Fe$_{40}$Co$_{30}$Cr$_{30}$Mn$_{5}$Ni$_{5}$
    \item Fe$_{20}$Co$_{40}$Cr$_{30}$Mn$_{5}$Ni$_{5}$
    \item Fe$_{5}$Co$_{20}$Cr$_{40}$Mn$_{5}$Ni$_{30}$.
\end{enumerate}
For the printability analysis, a 17-by-14 design of experiments for power and velocity was used. The parameters that were kept constant include the hatch spacing, the laser beam diameter, and the powder layer thickness at 70 $\mu$m, 80 $\mu$m, and 30 $\mu$m, as it was for the equiatomic CoCrFeMnNi alloy. 

\begin{figure*}
    \centering
    \includegraphics[width=1\textwidth]{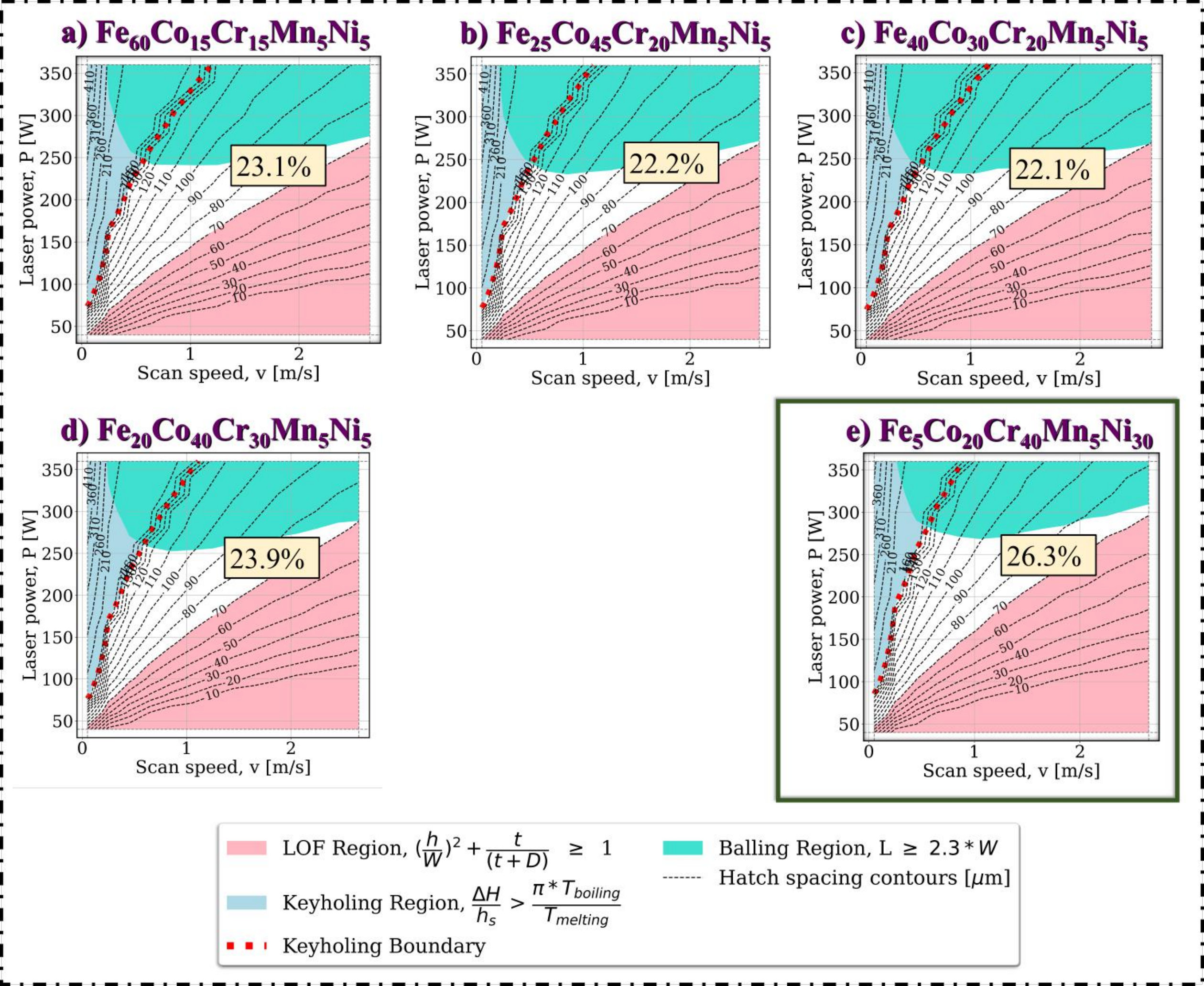}
    \caption{Using the analytical E-T thermal model, the printability maps for the five sampled alloys were generated with their respective printability index. The alloy with the highest printability of 26.3\% was Fe$_{5}$Co$_{20}$Cr$_{40}$Mn$_{5}$Ni$_{30}$ while the minimum printability index at 22.1\%. However, both values are greater than the printability index of the printability map of the equiatomic CoCrFeMnNi generated using the analytical E-T model.}
    \label{fig:Analytical_pareto}
\end{figure*}

Using the analytical thermal E-T model, the melt pool profile for each of the sampled alloys was obtained, and printability maps were constructed, as shown in Figure \ref{fig:Analytical_pareto}. The printability index can be used to rank the alloys with the maximum printability index to the lowest as the following: 
\begin{enumerate}
    \item Fe$_{5}$Co$_{20}$Cr$_{40}$Mn$_{5}$Ni$_{30}$
    \item Fe$_{20}$Co$_{40}$Cr$_{30}$Mn$_{5}$Ni$_{5}$
    \item Fe$_{60}$Co$_{15}$Cr$_{15}$Mn$_{5}$Ni$_{5}$
    \item Fe$_{25}$Co$_{45}$Cr$_{20}$Mn$_{5}$Ni$_{5}$
    \item Fe$_{40}$Co$_{30}$Cr$_{30}$Mn$_{5}$Ni$_{5}$.
\end{enumerate}

\begin{figure*}
    \includegraphics[width=1\textwidth]{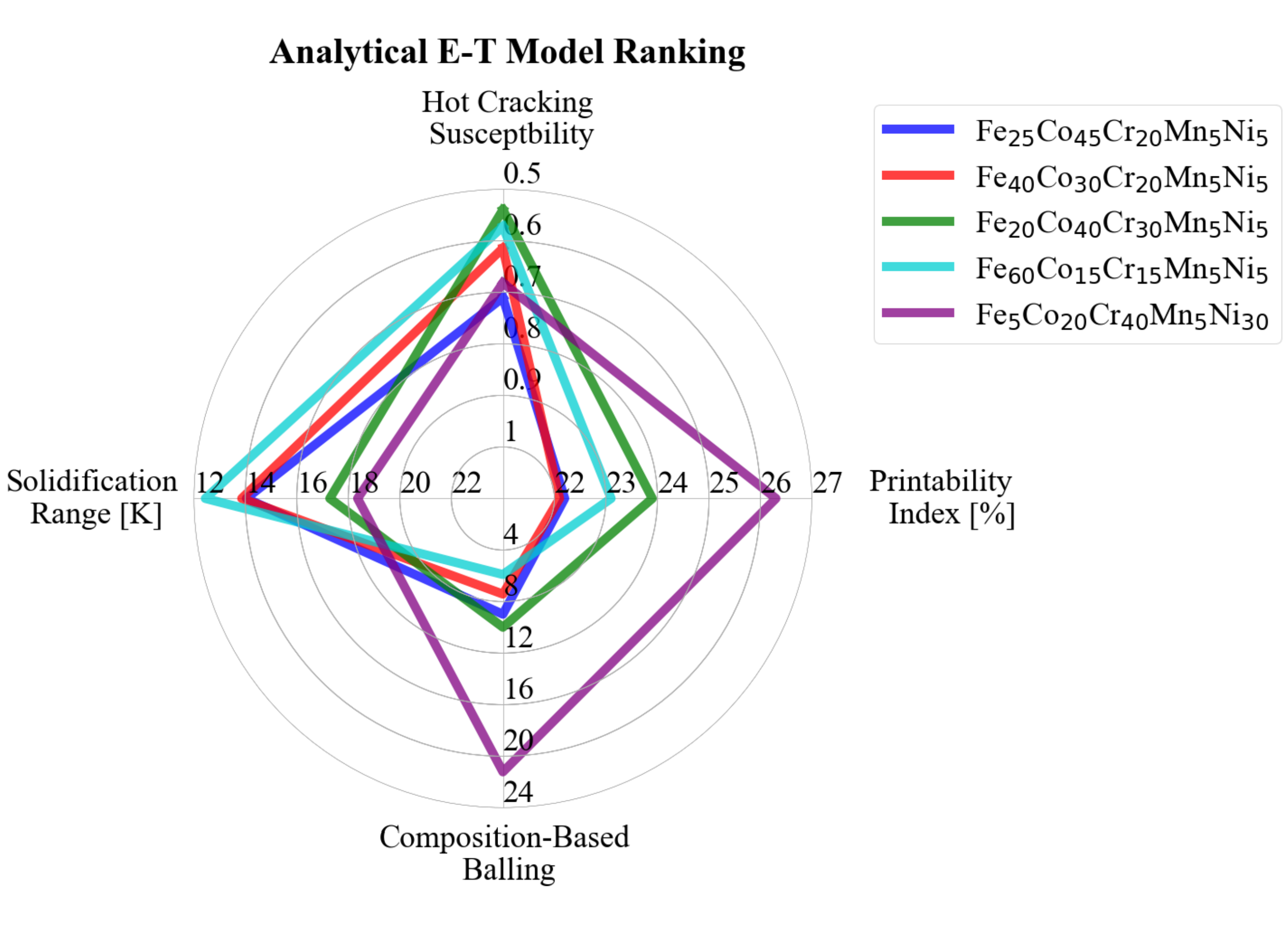}
    \caption{The radar chart summarizes the objectives used to rank the alloys for printability when the analytical E-T model is used to calculate the melt pool geometry. The objectives that are maximized are the printability index and the composition-based balling criterion, while the solidification range and the hot cracking susceptibility are minimized. To account for the two different optimization problems, the axis for the solidification range and hot cracking susceptibility is inverted such that the radar chart maximizes all objectives.}
    \label{fig:Ranking_Ana}
\end{figure*}

Furthermore, the ranking of each alloy can further be visualized  in Figure \ref{fig:Ranking_Ana} based on the objectives: hot cracking susceptibility, solidification range, composition-based balling criteria, and the printability index. The indices that are maximized are the printability index and composition-based balling, while the hot cracking susceptibility and the solidification range are minimized. In Figure \ref{fig:Ranking_Ana}, each objective is maximized where the axis for hot cracking susceptibility and solidification range is inverted.

\begin{figure*}
    \centering
    \includegraphics[width=1\textwidth]{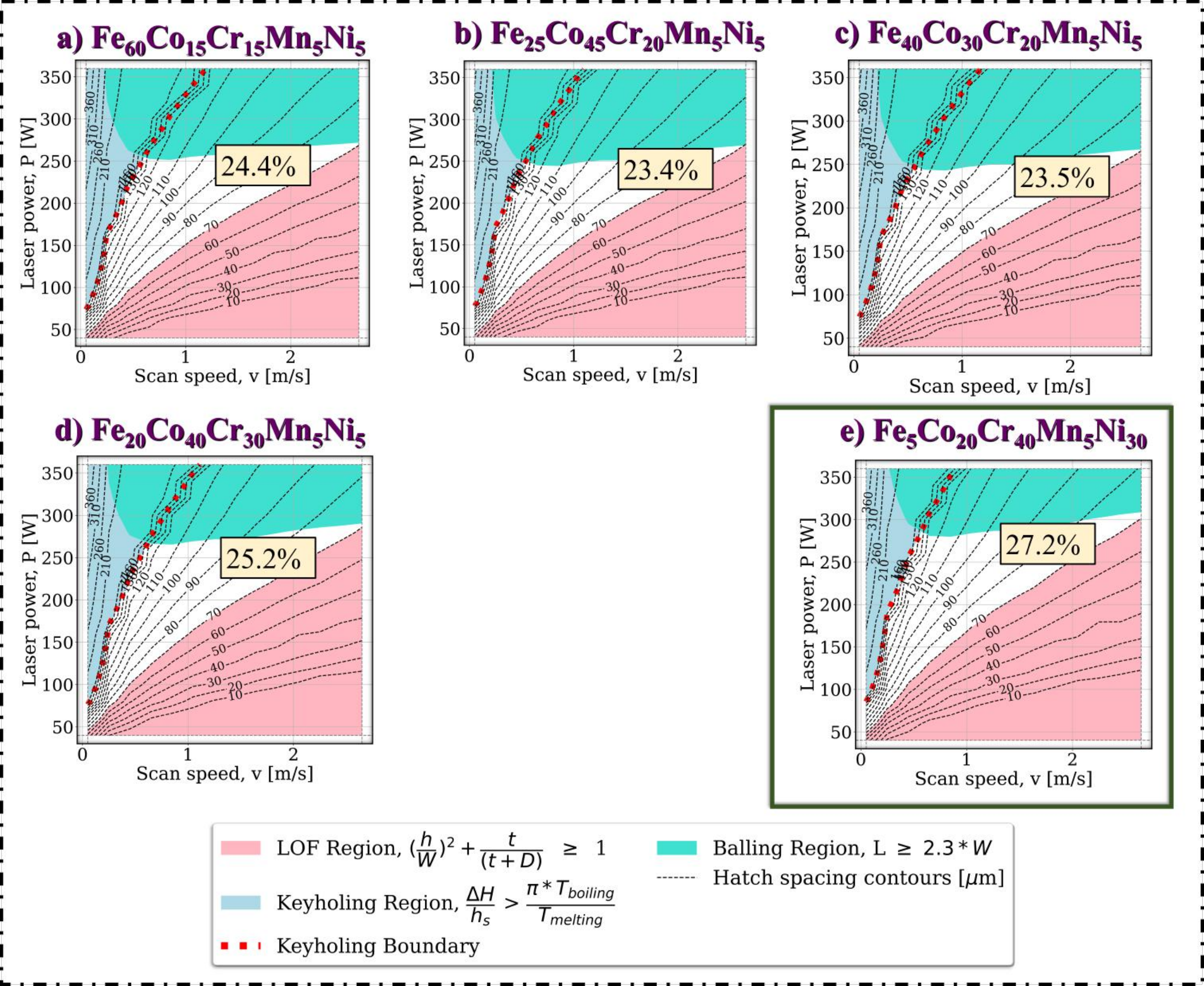}
    \caption{The melt pool profile for the five sampled alloys was obtained using the NN E-T model with the highest printability index of 27.2\% for Fe$_{5}$Co$_{20}$Cr$_{40}$Mn$_{5}$Ni$_{30}$. The alloy with the lowest printability index was for the Fe$_{60}$Co$_{15}$Cr$_{15}$Mn$_{5}$Ni$_{5}$ at 24.4\%. Both the values of the printability index are greater than the printability index of 18.5\% obtained from the printability map of equiatomic CoCrFeMnNi using the NN E-T model.}
    \label{fig:NN_ET_Pareto}
\end{figure*}

The maximum printability index had a value of 26.3\%, which is greater than the printability index determined for the equiatomic CoCrFeMnNi alloy printability map obtained using the analytical E-T thermal model of 17.1\%. The NN-ET thermal model had a higher maximum printability index of 27.2\% in comparison to the analytical ET thermal model, as shown in Figure \ref{fig:NN_ET_Pareto}. In addition, the current printability index is also larger than 18.5\%, which was obtained for the printability maps constructed for the equiatomic CoCrFeMnNi alloy using the NN E-T thermal model.  When ranking the alloys based on the printability index, the ranking is similar to that of the printability maps generated using the analytical E-T thermal model. The ranking for the printability maps for the sampled alloys is the following: 
\begin{enumerate}
    \item Fe$_{5}$Co$_{20}$Cr$_{40}$Mn$_{5}$Ni$_{30}$ 
    \item Fe$_{20}$Co$_{40}$Cr$_{30}$Mn$_{5}$Ni$_{5}$
    \item Fe$_{60}$Co$_{15}$Cr$_{15}$Mn$_{5}$Ni$_{5}$ 
    \item Fe$_{40}$Co$_{30}$Cr$_{30}$Mn$_{5}$Ni$_{5}$
    \item Fe$_{25}$Co$_{45}$Cr$_{20}$Mn$_{5}$Ni$_{5}$.
\end{enumerate}

\begin{figure*}
    \includegraphics[width=1\textwidth]{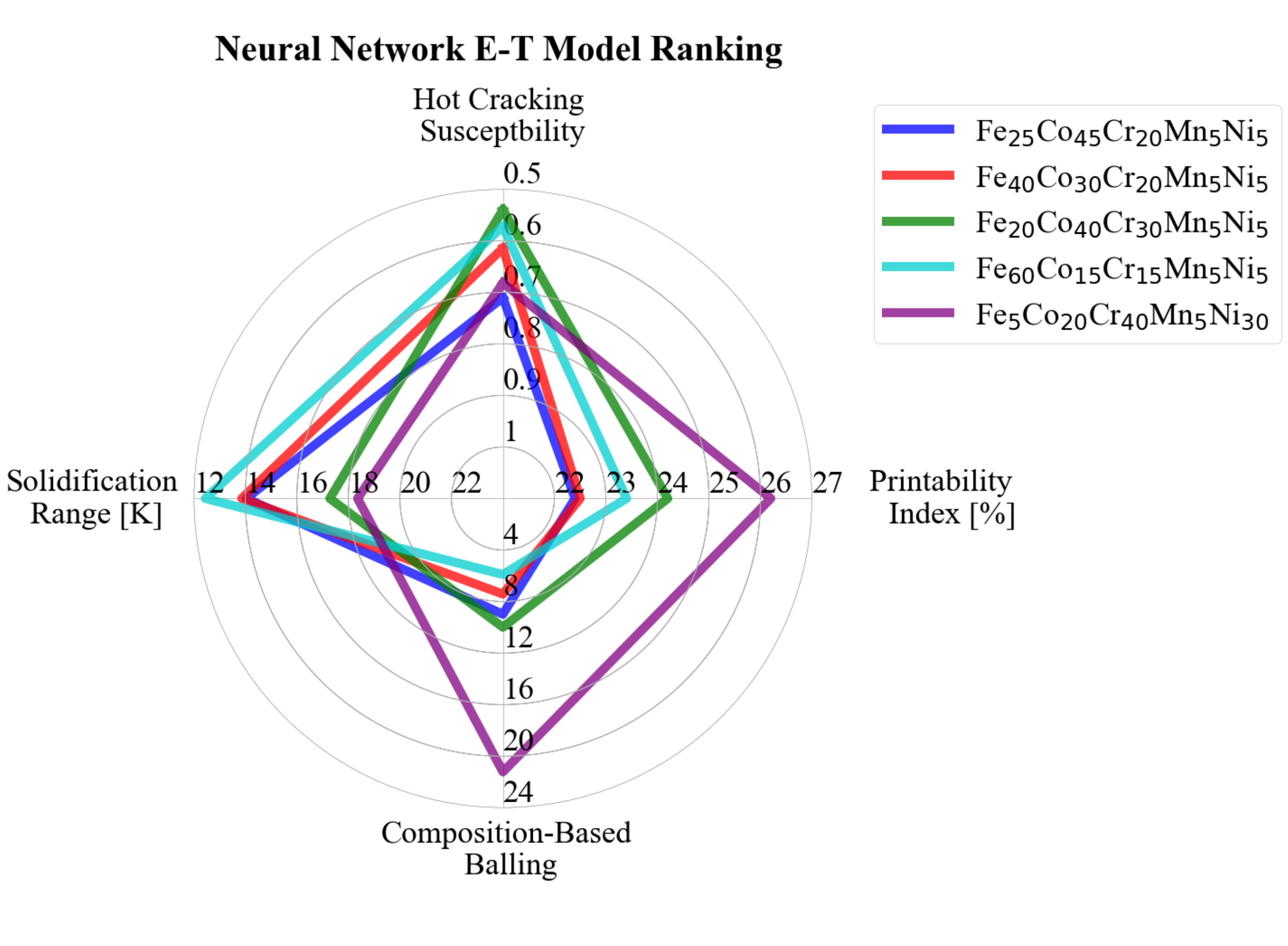}
    \caption{The radar chart summarizes the results for the alloys predicted using the pareto front objectives and further helps analyze the performance and printability of the alloys based on the four objectives: hot cracking susceptibility, solidification range, composition-based balling and the printability index when using the neural network E-T model. The axis for hot cracking susceptibility and solidification range is inverted to maximize all of the objectives.}
    \label{fig:Ranking_NN}
\end{figure*}

The ranking of the alloys when using the neural network E-T model is summarized graphically in Figure \ref{fig:Ranking_NN}. As done for Figure \ref{fig:Ranking_Ana}, the axis for hot cracking susceptibility and solidification range was inverted to create a maximization problem that the radar chart covers. 

We would like to note that the ranking between printability maps ranked at 3 and 4 has a difference in printable indices of 0.1\% for both thermal models.

\begin{figure*}
    \centering
    \includegraphics[width=1\textwidth]{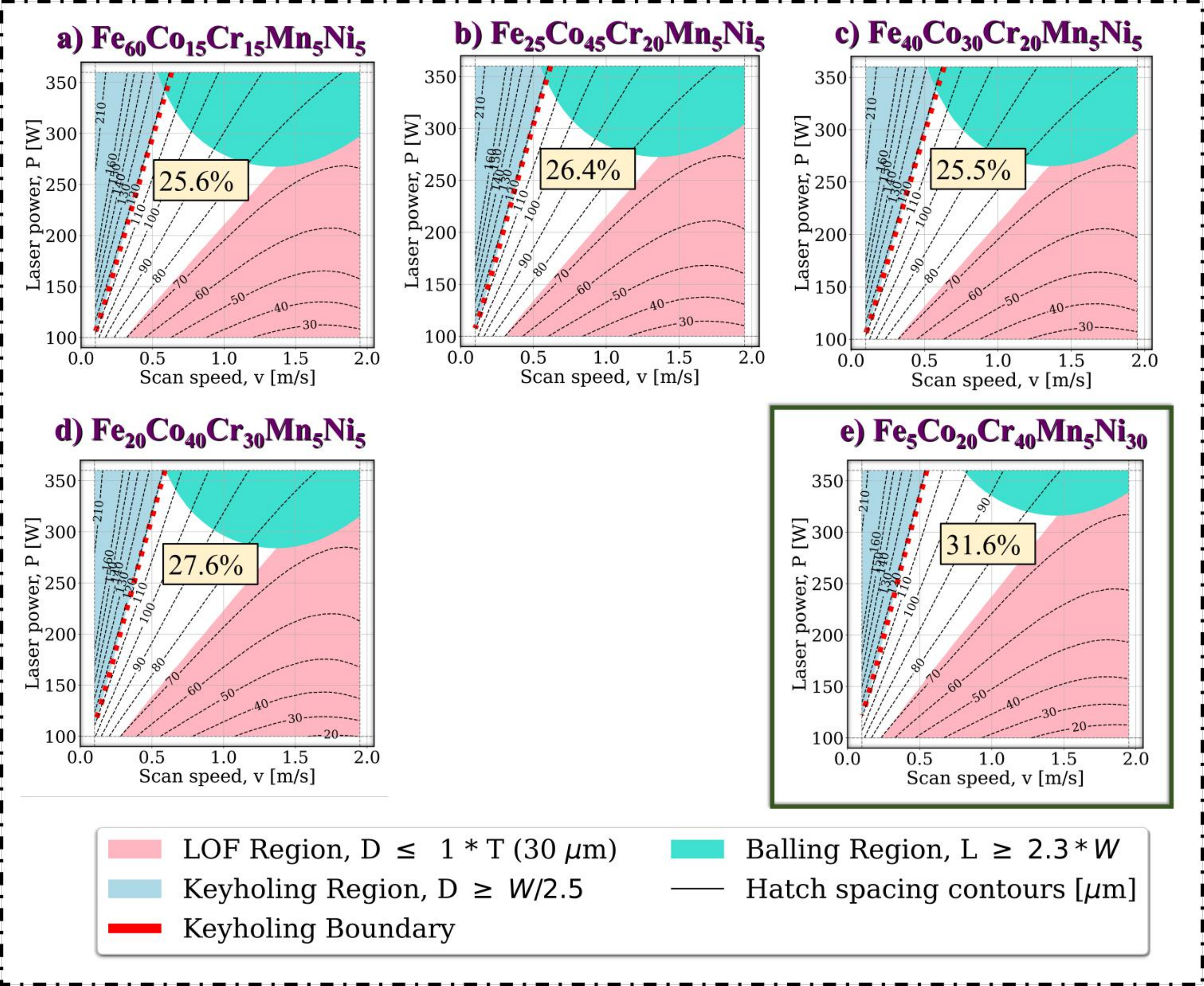}
    \caption{With the criteria set that gave the highest accuracy for the equiatomic CoCrFeMnNi printability maps fabricated using the melt pool profile obtained from dimensionless E-T thermal model, the maximum printability index obtained was 31.6\% for Fe$_{5}$Co$_{20}$Cr$_{40}$Mn$_{5}$Ni$_{30}$ and the minimum value for the printability index was 25.5\%. The values are both greater than the printability index for the printability maps for the equiatomic CoCrFeMnNi using the dimensionless E-T model of 22.2\%. }
    \label{fig:Scaled_ET_pareto_ideal}
\end{figure*}

\begin{figure*}
    \centering
    \includegraphics[width=1\textwidth]{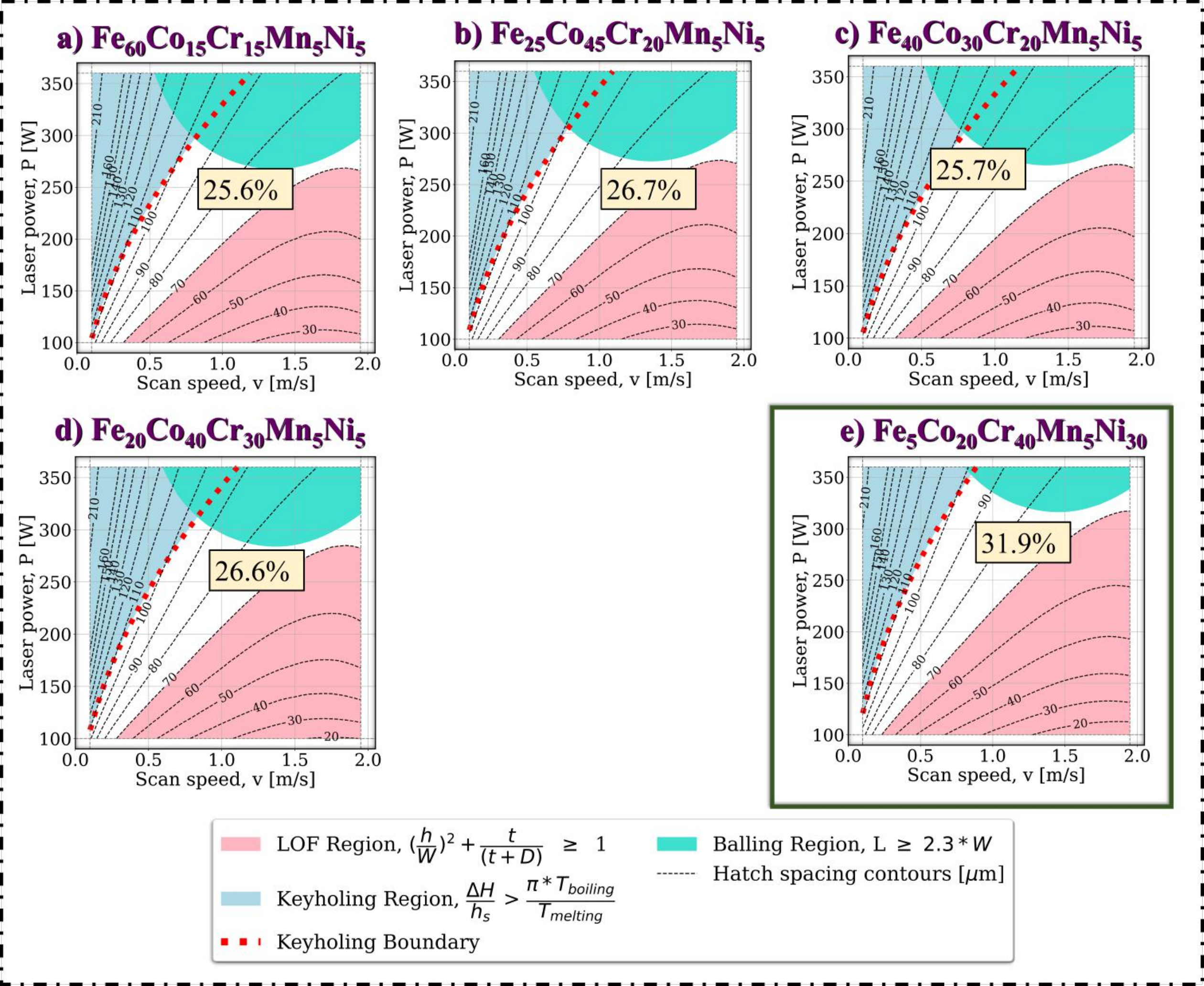}
    \caption{Printability maps with the optimal combination criteria set for the five sampled alloys were fabricated using the dimensionless E-T model. The maximum printability index obtained was 31.9\% for Fe$_{5}$Co$_{20}$Cr$_{40}$Mn$_{5}$Ni$_{30}$ and the minimum value for the printability index was 25.6\%. The values are both greater than the printability index for the printability maps for the equiatomic CoCrFeMnNi using the dimensionless E-T model of 22.2\%.}
    \label{fig:Scaled_ET_pareto_common}
\end{figure*}

For the dimensionless E-T thermal model and the general optimal criteria set of Equation \ref{lof2} for lack-of-fusion, Equation \ref{key2} for keyholing and Equation \ref{ball1}, the maximum printability index obtained was 31.9\% for the Fe$_{5}$Co$_{20}$Cr$_{40}$Mn$_{5}$Ni$_{30}$ alloy, as shown in Figure \ref{fig:Scaled_ET_pareto_common}. To compare, the printability maps for the sampled alloys were generated using the combination criteria set that uses Equation \ref{lof1} for lack-of-fusion, Equation \ref{key1} for keyholing, and Equation \ref{ball1} for balling, as shown in Figure \ref{fig:Scaled_ET_pareto_ideal}.

These criteria set was the combination of porosity-defect that evaluated the highest accuracy for the equiatomic CoCrFeMnNi alloy when the printability map was generated using the dimensionless E-T model. The maximum printability index for the set of printability maps was 31.6\%, which is greater than the printability index of 22.2\% that was evaluated for the equiatomic CoCrFeMnNi alloy printability map generated using the dimensionless E-T model. For the general optimal criteria set, the ranking of the printability maps generated using the dimensionless E-T model is the following: 
\begin{enumerate}
    \item Fe$_{5}$Co$_{20}$Cr$_{40}$Mn$_{5}$Ni$_{30}$ 
    \item Fe$_{25}$Co$_{45}$Cr$_{20}$Mn$_{5}$Ni$_{5}$ 
    \item Fe$_{20}$Co$_{40}$Cr$_{30}$Mn$_{5}$Ni$_{5}$ 
    \item Fe$_{40}$Co$_{30}$Cr$_{30}$Mn$_{5}$Ni$_{5}$
    \item Fe$_{60}$Co$_{15}$Cr$_{15}$Mn$_{5}$Ni$_{5}$.
\end{enumerate}

\begin{figure*}
    \includegraphics[width=1\textwidth]{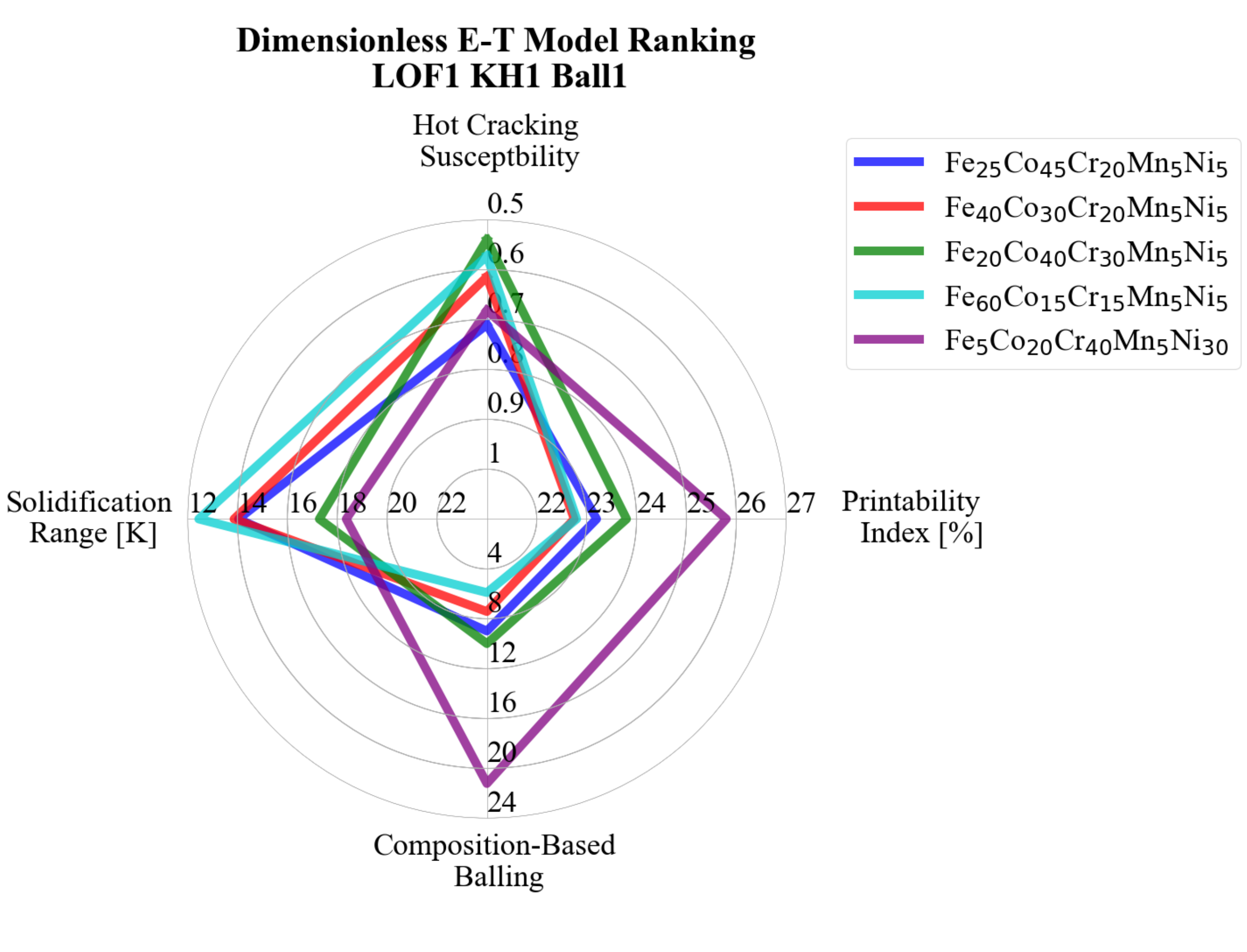}
    \caption{The radar chart displays the ranking of each alloy based on the Pareto objectives as well as the printability index. The axis for solidification range and hot cracking susceptibility is inverted to show the rankings where all objectives are being maximized.}
    \label{fig:Ranking_Dim1}
\end{figure*}

The ranking for the alloys based on the Pareto front objectives and printability index when the melt pool profile is predicted using the dimensionless form of the E-T model is summarized in Figure \ref{fig:Ranking_Dim1}.

On the other hand, for the other criteria set, the ranking of the printability maps is: 

\begin{enumerate}
    \item Fe$_{5}$Co$_{20}$Cr$_{40}$Mn$_{5}$Ni$_{30}$ 
    \item Fe$_{20}$Co$_{40}$Cr$_{30}$Mn$_{5}$Ni$_{5}$ 
    \item Fe$_{25}$Co$_{45}$Cr$_{20}$Mn$_{5}$Ni$_{5}$ 
    \item Fe$_{60}$Co$_{15}$Cr$_{15}$Mn$_{5}$Ni$_{5}$ 
    \item Fe$_{40}$Co$_{30}$Cr$_{30}$Mn$_{5}$Ni$_{5}$.  
\end{enumerate}

\begin{figure*}
    \includegraphics[width=1\textwidth]{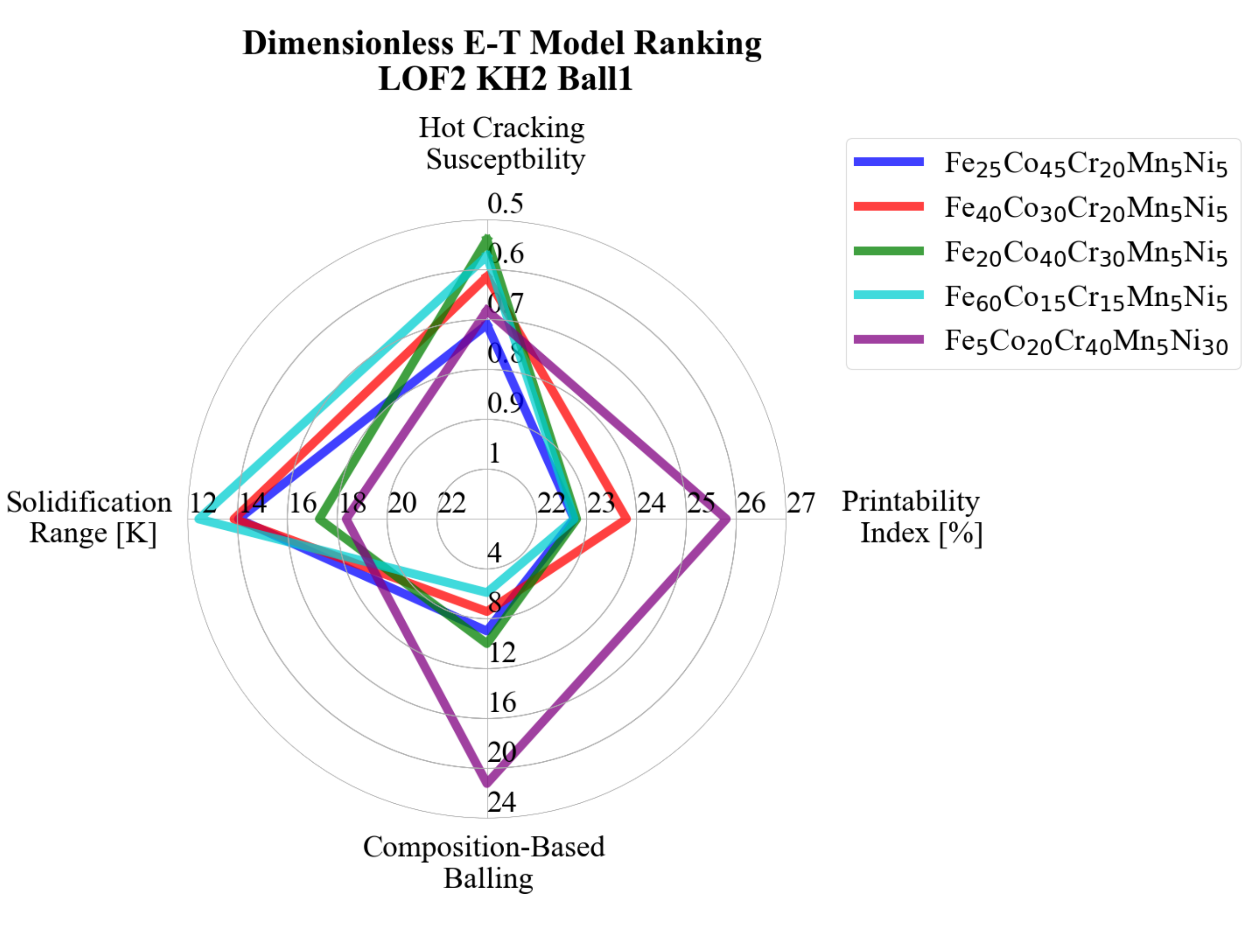}
    \caption{The radar chart graphically displays the ranking of the alloys based on the Pareto front objectives and printability index when using the dimensionless E-T model as well as the optimal combination criteria set. The axis for solidification range and hot cracking susceptibility is inverted to show the rankings where all objectives are being maximized.}
    \label{fig:Ranking_Dim2}
\end{figure*}

The ranking for the alloys based on the Pareto front objectives and printability index when the melt pool profile is predicted using the dimensionless form of the E-T model using the optimal combination criteria set is displayed in Figure \ref{fig:Ranking_Dim2}.

The ranking for the dimensionless E-T model is slightly different from the rankings of the printability maps obtained from the previous two thermal models. However, there is a consensus, as the alloy that has the highest printability is alloy Fe$_{5}$Co$_{20}$Cr$_{40}$Mn$_{5}$Ni$_{30}$ for all three thermal models. 

By comparing the printability index from the equiatomic CoCrFeMnNi and the five sampled alloys, we can evaluate that our proposed framework allows us to design a HEA that contains elements Co,Cr,Fe,Mn,Ni that can be printed in a large power and velocity processing space.  

\subsection{Opportunities for High-Throughput Evaluation of Printability and Alloy Design Opportunities}

\begin{figure*}[htb!]
    \centering
    \includegraphics[width=.7\textwidth]{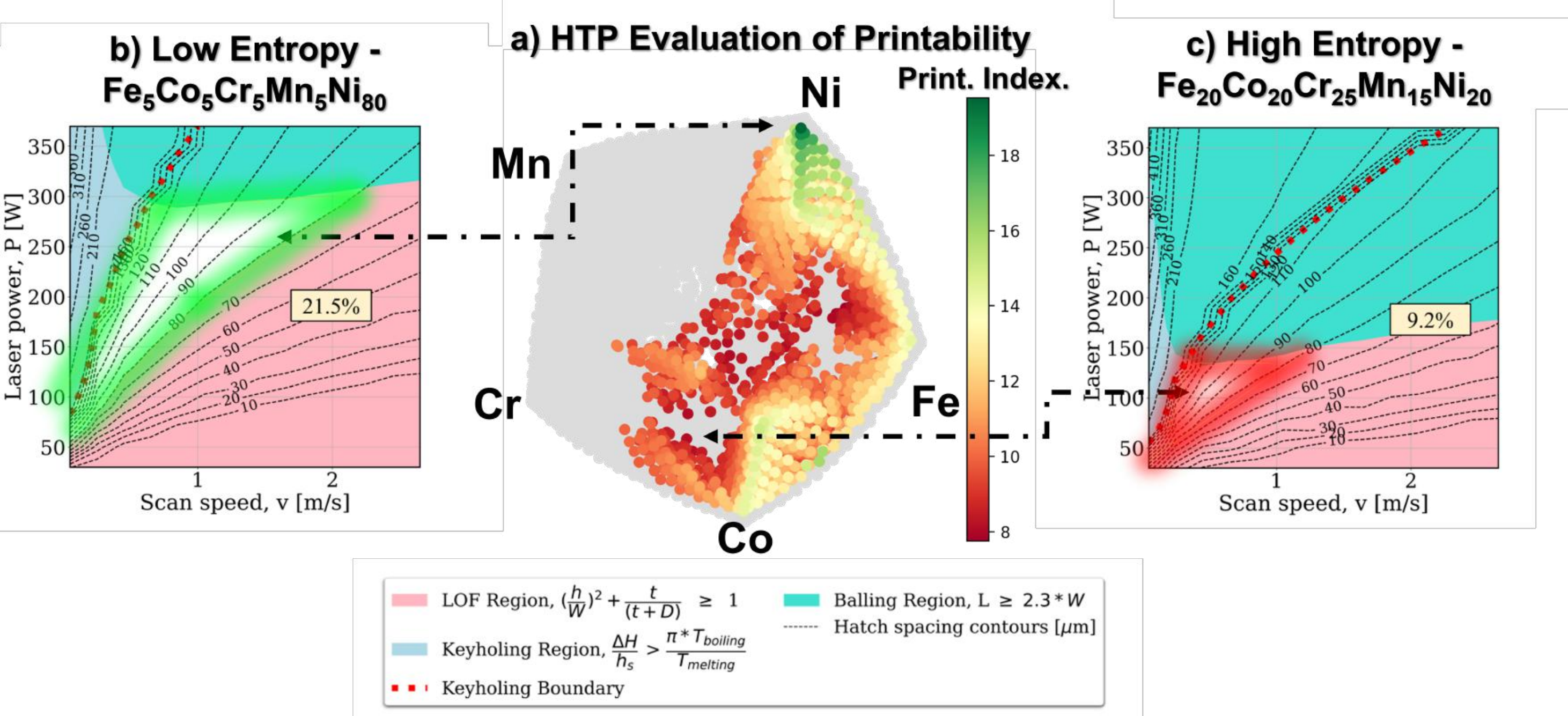}
    \caption{a) UMAP projection of the design space. The printability index is plotted over this space. By comparing a c) high entropy alloy and a  b) low entropy alloy printability maps, it can be observed that the low entropy alloy has a high printability index as defects such as balling and keyholing are less prominent in comparison to the high entropy alloy. The printability index for the low entropy alloy is 21.5\%. On the other hand, the high entropy alloy has a printability index of 9.2\%.}
    \label{fig:HivsLow}
\end{figure*}

In the previous analysis, 5 alloys were designed with composition-based printability objectives that aim to minimize balling and solidification cracking. \emph{However, there exist opportunities to tailor alloys such that they have optimal printability maps.} From our previous analysis, 1,610 alloys satisfied the 1300$^o$C single-phase FCC constraint. Using the ET-NN model and the optimal criteria combination determined in Section \ref{sec:Arb_cantor}, the intrinsic printability of the 1,610 alloys was evaluated in a HTP manner. Specifically, the printability maps for the 1,610 constraint-satisfying alloys were queried, and the printability index was extracted from each printability map. As defined above, the printability index is defined as the amount of area that is defect free in a given processing range. Using this index as an alloy design metric, alloys can be ranked according to their processable windows. When queried for the 1,610 feasible alloys, shown in Fig. \ref{fig:HivsLow}, it is evident that Ni-rich alloys are the \emph{most intrinsically printable alloys} within the Co-Cr-Fe-Mn-Ni alloy space. Another feature that is evident from Fig. \ref{fig:HivsLow} is that alloys with \emph{higher nominal entropy}, i.e., those in the central regions of the composition space, \emph{are less printable}.

The most prominent difference between the high entropy and low entropy alloys is the size of the balling defect region in the printability maps, as shown in Fig. \ref{fig:HivsLow} b and c. The criterion used to determine balling in this printability map is defined in Equation \ref{ball1}, which is based on the ratio between melt pool length and melt pool width. The difference in L/W ratios between these two alloys can be explained by the fact that increasing alloy complexity increases phonon and electron scattering which in turn decreases the thermal conductivity of the alloy \cite{CHOU2009184}. The Cantor alloy with a higher entropy has a predicted thermal conductivity of 21 $W/m/K$ while the Cantor alloy with a lower entropy has a predicted thermal conductivity of 38 $W/m/K$. The thermal conductivity has a direct impact on the melt pool length. As the thermal conductivity decreases, the time required for the melt pool to solidify increases ($4.66\times10^{-4}$ s for the Cantor alloy with a lower entropy and $6.58\times 10^{-4}$ s for the Cantor alloy with a higher entropy, as predicted with Equation \ref{solid_tau2}), which in turn results in a long melt pool of  $4.38\times 10^{-4}$ m for the Cantor alloy with higher entropy and  $2.58\times 10^{-4}$ for the Cantor alloy with lower entropy (given P = 250 W and v = 1.45 $W/m/K$). This long liquid melt pool is more susceptible to Rayleigh capillary instability. While the width is affected by the thermal conductivity, it is not affected to the same extent as the melt pool length. Therefore (given P = 250 W and v = 1.45 $m/s$), the L/W ratio of the Cantor alloy with higher entropy (L/W = 3.65) is greater than that of the L/W ratio for the Cantor alloy with lower entropy (L/W = 2.09), resulting increased susceptibility to balling for the Cantor alloy with the higher entropy.

Alloy design requires \emph{optimizable} indicators of quantities of interest associated with the alloy. Alloy design for amenability to AM requires easily queriable metrics that provide an indication of the size of the window of processing parameters that result in defect-free prints. Just as Vela et al. \cite{vela2022evaluating} demonstrated an indicator that is correlated with the extent of the processing window that results in balling, the printability index can be \emph{maximized} in alloy design schemes, yielding alloys that have a wide range of processing parameters that result in good prints. If combined with printability indicators that describe the tendency toward hot-cracking (i.e., solidification range and CSC), there are opportunities to design \emph{intrinsically printable alloys}, i.e., alloys resistant to processing-related porosity and solidification-related cracking.

\section{Conclusion}

The methodology proposed in the current work introduces a physics-based computational framework that can aid in designing an alloy system for high printability in a computationally inexpensive manner. Despite the vast research conducted in AM, the design for AM is expensive due to the large material composition and processing space. Most designs already proposed in the literature for AM consist of  trial-and-error,  as well as high-fidelity simulations that help evaluate optimized processing parameters. In addition to optimizing processing space, it can be recognized that material compositions are more highly susceptible to material intrinsic defects such as hot cracking than other compositions. Therefore, to accelerate the design for L-PBF, a framework is required that can reduce the composition space and processing space that will lead to a material system at optimized processing conditions for defect-free AM products. 

Various criteria are introduced in the current work for composition-informed defects such as hot cracking and balling as well as process-induced defects, specifically lack-of-fusion, keyholing, and balling. To evaluate for hot cracking, a hot cracking susceptibility coefficient proposed by Clyne and Davies is used to narrow down the composition space for Co-Cr-Fe-Mn-Ni HEA systems. In addition, a composition-informed balling susceptibility metric that has been validated against over 2,000 unique alloy-process observations and over 155 unique compositions was used to further narrow down the composition space where the intrinsic material property is not heavily influenced by balling. On the other hand, two criteria for lack-of-fusion, three criteria for keyholing, and two criteria for balling are used to identify regions of the respective defects in the processing space. The process-induced defects are a function of material properties, melt pool profile, and/or processing parameters.

The processing space with defined regions of the lack-of-fusion, keyholing, and balling can be visualized using a printability map, a 2D visualization in the processing space can identify porosity-free regions as well as regions in the processing space where the process-induced defects do occur. However, before the printability maps can be constructed using a combination of process-induced defects, the material properties are needed to evaluate the melt pool profile in the processing space. The material properties are calculated using a CALPHAD model using the Thermo-Calc software, and the melt pool profile is evaluated using one of the three E-T models, the analytical, NN, and dimensionless E-T thermal models. However, the analytical thermal model can be computationally expensive as a grid of 200 points in the power and velocity space can take 1.5 hours, while for the NN E-T model and the dimensionless E-T model, the same grid of power and velocity point would take only 3 to 5 seconds. In addition, with the material properties calculated with the CALPHAD model, we can also remove the calibration of the E-T model parameters, which increases the expense of constructing printability maps to about 10 days. The use of the E-T ML surrogate is the enabling tool that made the HTP assessment of printability possible.

To evaluate our framework, we proposed designing a HEA system consisting of elements Co-Cr-Fe-Mn-Ni with a single FCC phase structure for high-temperature applications. The Co-Cr-Fe-Mn-Ni alloy system was chosen as it is one of the most widely researched HEA systems, specifically the equiatomic CoCrFeMnNi alloy (i.e., Cantor alloy system). To determine the optimal process-induced criteria set, the twelve printability maps for the Cantor alloy system were constructed using the three E-T models for a total of 36 printability maps. Experimental points mined from literature in conjunction with experiments conducted in-house were overlaid on the printability maps to find the highest accuracy to determine the optimal combination of criteria for lack-of-fusion, keyholing, and balling. For the analytical E-T model and the NN E-T model, the highest accuracy was 76\%. The printability map with the highest accuracy was constructed using Equation \ref{lof2} as the criterion for lack-of-fusion, Equation \ref{key2} for the keyholing criterion, and Equation \ref{ball1} for the criterion used to evaluate the balling region. For the dimensionless E-T model, the printability with the highest accuracy of 72\% was constructed using Equation \ref{lof1} for the lack-of-fusion criterion, Equation \ref{key1} for the criterion used to evaluate keyholing region and Equation \ref{ball1} for the balling criterion. Based on the equiatomic CoCrFeMnNi alloy, we can determine that the optimized set of criteria for a Co-Cr-Fe-Mn-Ni HEA system is Equation \ref{lof2}, Equation \ref{key2}, and Equation \ref{ball1}.

To further design the HEA material for a high-temperature application, a tri-objective Pareto front was used to screen for alloys that are predicted to have a single FCC phase structure at $1300^\circ C$ using a CALPHAD model, low hot cracking susceptibility, and low composition-informed balling susceptibility. Using a k-medoids sampling method five different compositions were chosen to be analyzed with our proposed framework. Printability maps using the optimized criteria set were used to fabricate printability maps for alloys using all three thermal models:
\begin{enumerate}
    \item Fe$_{60}$Co$_{15}$Cr$_{15}$Mn$_{5}$Ni$_{5}$;
    \item Fe$_{25}$Co$_{45}$Cr$_{20}$Mn$_{5}$Ni$_{5}$; 
    \item Fe$_{40}$Co$_{30}$Cr$_{30}$Mn$_{5}$Ni$_{5}$; 
    \item Fe$_{20}$Co$_{40}$Cr$_{30}$Mn$_{5}$Ni$_{5}$; 
    \item Fe$_{5}$Co$_{20}$Cr$_{40}$Mn$_{5}$Ni$_{30}$.
\end{enumerate}
 The area of the defect-free region also called the printability index was evaluated and used to rank the printability maps for the material that is most printable in the processing space. All three models were in agreement that Fe$_{5}$Co$_{20}$Cr$_{40}$Mn$_{5}$Ni$_{30}$ was the alloy that was most printable with a printability index of 26.3\%, 27.2\% and 33.9\% for the maps generated using the analytical E-T model, the NN E-T model and dimensionless form of the E-T model, respectively. The printability index was larger for these five alloys in comparison to the printability index for the equiatomic CoCrFeMnNi alloy. This highlights the effectiveness of our framework to design a HEA system for high-temperature applications as well as constructing printability maps in a computationally inexpensive manner. In conclusion, the authors have introduced a fully computational framework that can help expedite L-PBF product design. This approach can be used to explore vast alloy spaces in order to carry out HTP-based approaches to alloy design in the context of AM.

\section*{Acknowledgements}
We acknowledge Dr. Bing Zhang and Raiyan Seede for performing several experiments captured in the database used in this work, under Army Research Office (ARO) Contract No. W911NF-18-1-0278. PH and RA also acknowledge the support of NSF, United States, through Grant No. 1849085. BV, SS, and DS acknowledge Grant no. NSF-DGE-1545403 (NSF-NRT: Data-Enabled Discovery and Design of Energy Materials, D$^3$EM). The authors would also like to acknowledge the NASA-ESI Program under Grant Number 80NSSC21K0223, as well as ARPA-E ULTIMATE Program through Project DE-AR0001427. High-throughput calculations were carried out in part at the Texas A\&M High-Performance Research Computing (HPRC) Facility.

\bibliographystyle{elsarticle-num}
\bibliography{Reference}

\end{document}